%% file: EXO-17-009_temp.tex
\pdfoutput=1

\documentclass[11pt,twoside,a4paper,cmspaper,final,collab]{cms-tdr}
\def\svnVersion{492099P}\def\cmsCernNoTag{CERN-EP-2018-265}\def\cmsCernDate{\today}\def\cmsMessage{Published in Physical Review D as \href{http://dx.doi.org/10.1103/PhysRevD.99.052002}{\doi{10.1103/PhysRevD.99.052002.}}}

\begin{document}\cmsNoteHeader{EXO-17-009}

\hyphenation{had-ron-i-za-tion}
\hyphenation{cal-or-i-me-ter}
\hyphenation{de-vices}
\RCS$Revision: 491905 $
\RCS$HeadURL: svn+ssh://svn.cern.ch/reps/tdr2/papers/EXO-17-009/trunk/EXO-17-009.tex $
\RCS$Id: EXO-17-009.tex 491905 2019-03-15 15:12:09Z scooper $

\ifthenelse{\boolean{cms@external}}{\providecommand{\cmsLeft}{upper\xspace}}{\providecommand{\cmsLeft}{left\xspace}}
\ifthenelse{\boolean{cms@external}}{\providecommand{\cmsRight}{lower\xspace}}{\providecommand{\cmsRight}{right\xspace}}
\ifthenelse{\boolean{cms@external}}{\providecommand{\cmsMiddle}{middle\xspace}}{\providecommand{\cmsMiddle}{lower left\xspace}}
\ifthenelse{\boolean{cms@external}}{\providecommand{\cmsRightThree}{lower\xspace}}{\providecommand{\cmsRightThree}{lower right\xspace}}
\ifthenelse{\boolean{cms@external}}{\providecommand{\NA}{\ensuremath{\cdots}\xspace}}{\providecommand{\NA}{\ensuremath{\text{---}}\xspace}}
\ifthenelse{\boolean{cms@external}}{\providecommand{\CL}{C.L.\xspace}}{\providecommand{\CL}{CL\xspace}}
\newcommand{\eejj}{\ensuremath{\Pe\Pe\text{jj}}\xspace}
\newcommand{\enujj}{\ensuremath{\Pe\Pgn\text{jj}}\xspace}
\newcommand{\mee}{\ensuremath{m_{\Pe\Pe}}\xspace}
\newcommand{\st}{\ensuremath{S_{\mathrm{T}}}\xspace}
\newcommand{\mej}{\ensuremath{m_{\Pe\text{j}}}\xspace}
\newcommand{\mejmin}{\ensuremath{m_{\Pe\text{j}}^\text{min}}\xspace}
\newcommand{\zjets}{\ensuremath{\cPZ/\PGg^{*}\text{+jets}}\xspace}
\newcommand{\wjets}{\ensuremath{\PW\text{+jets}}\xspace}
\newcommand{\vv}{\ensuremath{\text{VV}}\xspace}
\newcommand{\mlq}{\ensuremath{m_{\text{LQ}}}\xspace}
\newcommand{\gammajets}{\ensuremath{\PGg\text{+jets}}\xspace}
\ifthenelse{\boolean{cms@external}}{\providecommand{\cmsTable}[1]{#1}}{\providecommand{\cmsTable}[1]{\resizebox{\textwidth}{!}{#1}}}

\cmsNoteHeader{EXO-17-009}
\title{Search for pair production of first-generation scalar leptoquarks at \texorpdfstring{$\sqrt{s}=13\TeV$}{sqrt(s) = 13 TeV}}

\date{\today}

\abstract{
A search for the pair production of first-generation scalar leptoquarks is performed using proton-proton collision data recorded at 13\TeV center-of-mass energy with the CMS detector at the LHC. The data correspond to an integrated luminosity of 35.9\fbinv. The leptoquarks are assumed to decay promptly to a quark and either an electron or a neutrino, with branching fractions $\beta$ and $1-\beta$, respectively. The search targets the decay final states comprising two electrons, or one electron and large missing transverse momentum, along with two quarks that are detected as hadronic jets.  First-generation scalar leptoquarks with masses below 1435~(1270)\GeV are excluded for $\beta =\,1.0\,(0.5)$.  These are the most stringent limits on the mass of first-generation scalar leptoquarks to date.  The data are also interpreted to set exclusion limits in the context of an $R$-parity violating supersymmetric model, predicting promptly decaying top squarks with a similar dielectron final state.
}

\hypersetup{
pdfauthor={CMS Collaboration},
pdftitle={Search for pair production of first-generation scalar leptoquarks at sqrt(s) = 13 TeV},
pdfsubject={CMS},
pdfkeywords={CMS, physics, leptoquarks}}

\maketitle
\section{Introduction}
\label{sec:introduction}

The quark and lepton sectors of the standard model
(SM)~\cite{Glashow:1961tr,Weinberg:1967tq,Salam:1968rm} are similar:
both have the same number of generations composed of electroweak doublets.
This could indicate the existence of an additional
fundamental symmetry linking the two sectors, as proposed in
many scenarios of physics beyond the SM.  These include
grand unified theories with symmetry groups SU(4) of
the Pati--Salam model~\cite{gut0,gut1}, SU(5), SO(10),
and SU(15)~\cite{gut2,gut3,gut4,gut5,gut6,gut7};
technicolor~\cite{technicolor1,technicolor2,technicolor3};
superstring-inspired models~\cite{superstring_e6}; and models exhibiting
quark and lepton substructures~\cite{composite}.
A common feature of these models is the presence of a new class of bosons,
called leptoquarks (LQs), that carry both lepton ($L$) and baryon numbers
($B$). In general, LQs have fractional electric charge and are color
triplets under SU(3)$_\text{C}$. Their other properties, such as spin,
weak isospin, and fermion number ($3B+L$), are model dependent.

Direct searches for LQs at colliders are usually interpreted in the context
of effective theories that impose constraints on their interactions.
In order to ensure renormalizability, these interactions are required to
respect SM group symmetries, restricting the couplings of the LQs to SM
leptons and quarks only. A detailed account of LQs and their interactions
can be found in Ref.~\cite{mBRW}. Results from experiments sensitive to
lepton number violation, flavor changing neutral currents, and proton
decay allow the existence of three distinct generations of LQs with
negligible intergenerational mixing for mass scales accessible at the CERN
LHC~\cite{lq_constraints1,FCNC}. Indirect searches for new physics in rare \PB\ meson
decays~\cite{Aaij:2013qta,Aaij:2014ora,Aaij:2015oid,Aaij:2017vbb,Wehle:2016yoi}
by LHCb and Belle suggest a possible breakdown of lepton universality. These
anomalies, if confirmed, could provide additional support for LQ-based
models~\cite{Hiller:2018ijj}. A comprehensive review of LQ phenomenology and
experimental constraints on their properties is given in Ref.~\cite{Dorsner:2016wpm}.

We search for the pair production of first-generation scalar LQs that decay promptly.
The final state arising from each LQ decay comprises a quark
that is detected as a hadronic jet, and either an electron or
a large missing transverse momentum attributed to the presence
of an undetected neutrino. For light-quark final states,
the quark flavors cannot be determined from the observed jets.
We assume the LQs decay only to $\Pe\,(\Pgne)$ and up or down quarks.
The branching fractions for the LQ decay are expressed in terms of a
free parameter $\beta$, where $\beta$ denotes the branching fraction
to an electron and a quark, and $1-\beta$ the branching fraction to a
neutrino and a quark. For pair production of LQs, we consider two decay
modes. The first arises when each LQ decays to an electron and a quark,
having an overall branching fraction of $\beta^2$.  In the second mode
one LQ decays to an electron and a quark, and the other to a neutrino
and a quark. This mode has a branching fraction of $2\beta(1-\beta)$.
We, therefore, utilize final states with either two high transverse
momentum (\pt) electrons and two high-\pt\ jets, denoted as \eejj,
or one high-\pt\ electron, large missing transverse momentum,
and two high-\pt\ jets, denoted as \enujj.

Previous experiments at the LEP~\cite{Abbiendi:2003iv},
HERA~\cite{Chekanov:2003af,H1Collaboration:2011qaa}, and
Tevatron~\cite{Acosta:2005ge,Abazov:2009ab} colliders have searched for
LQ production and placed lower limits of several hundreds of~\GeV on
allowed LQ masses (\mlq) at 95\% confidence level (\CL). The CMS experiment
at the LHC has extended the limits on pair production of first-generation
scalar LQs using proton-proton (\Pp\Pp) collision data recorded during
2012 at a center-of-mass energy of $\sqrt{s} = 8\TeV$. Based on a sample
corresponding to an integrated luminosity of 19.7\fbinv, the lower limit
obtained on \mlq\ was 1010~(850)\GeV for $\beta=1.0\,(0.5)$~\cite{cms8tevPaper}.
The CMS Collaboration has also published results on a search
for singly produced LQs with the final states of either two
electrons and one jet, or two muons and one jet~\cite{cmsSingleLQ8tev}.
Recently, using a data set of 3.2\fbinv collected at $\sqrt{s}
= 13\TeV$, the ATLAS experiment has placed a lower limit on \mlq\ of
1100\GeV~\cite{atlasLQ2015paper} for $\beta = 1.0$.

\begin{figure*}[hbt]
   \centering
   {\includegraphics[width=.3\textwidth]{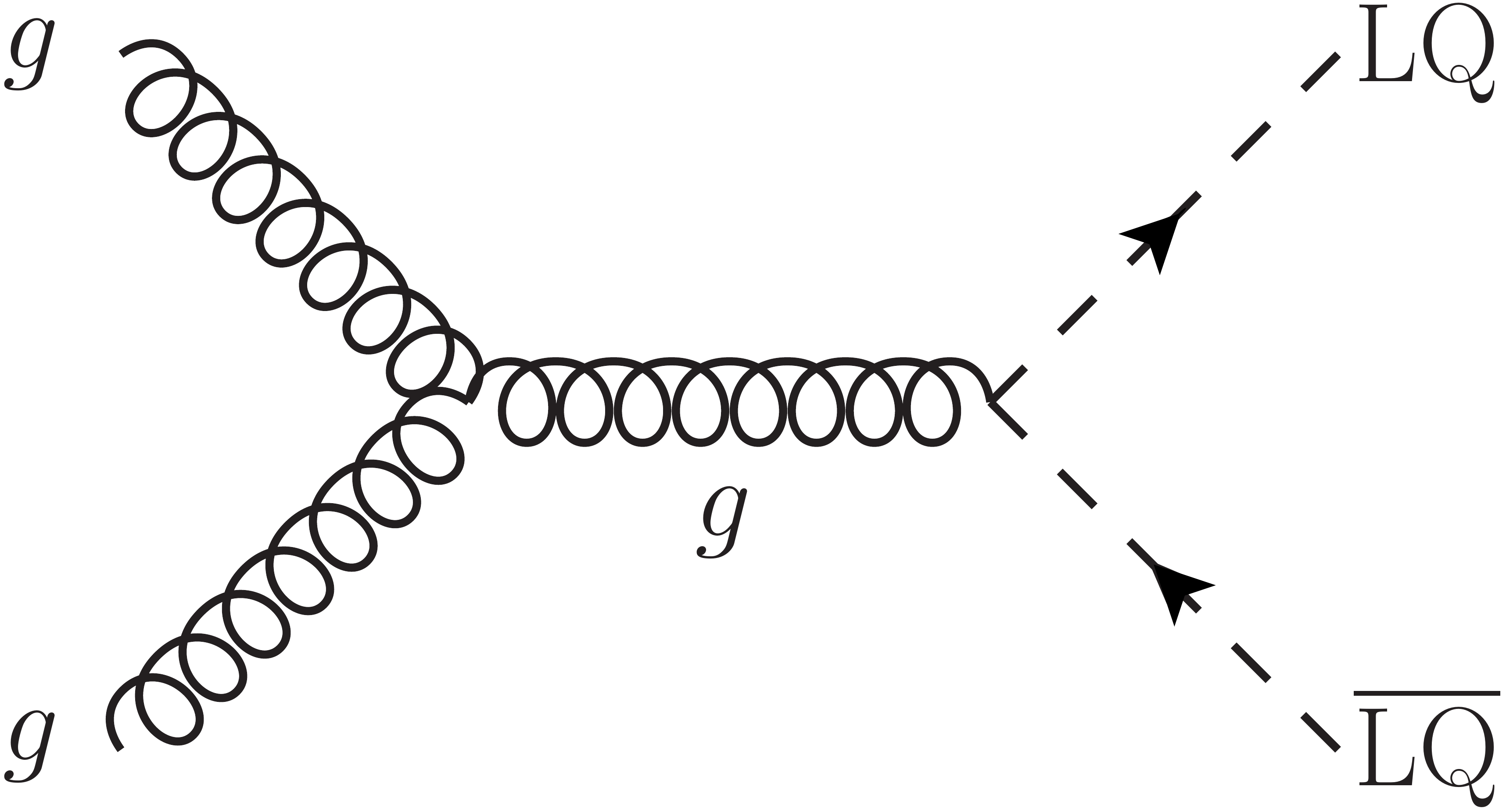}}\hspace*{0.4cm}
   {\includegraphics[width=.3\textwidth]{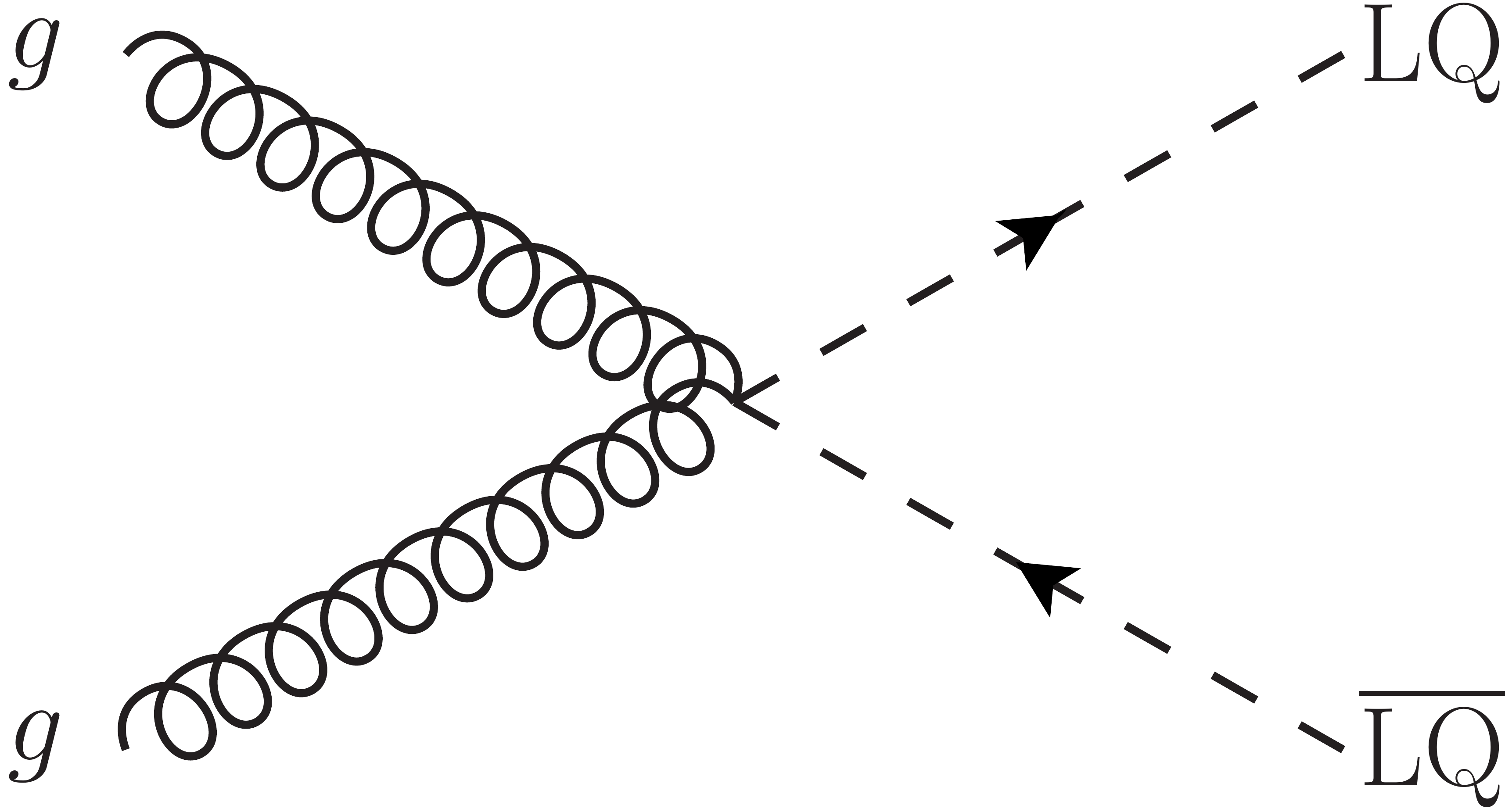}}\hspace*{0.4cm}
   {\includegraphics[width=.3\textwidth]{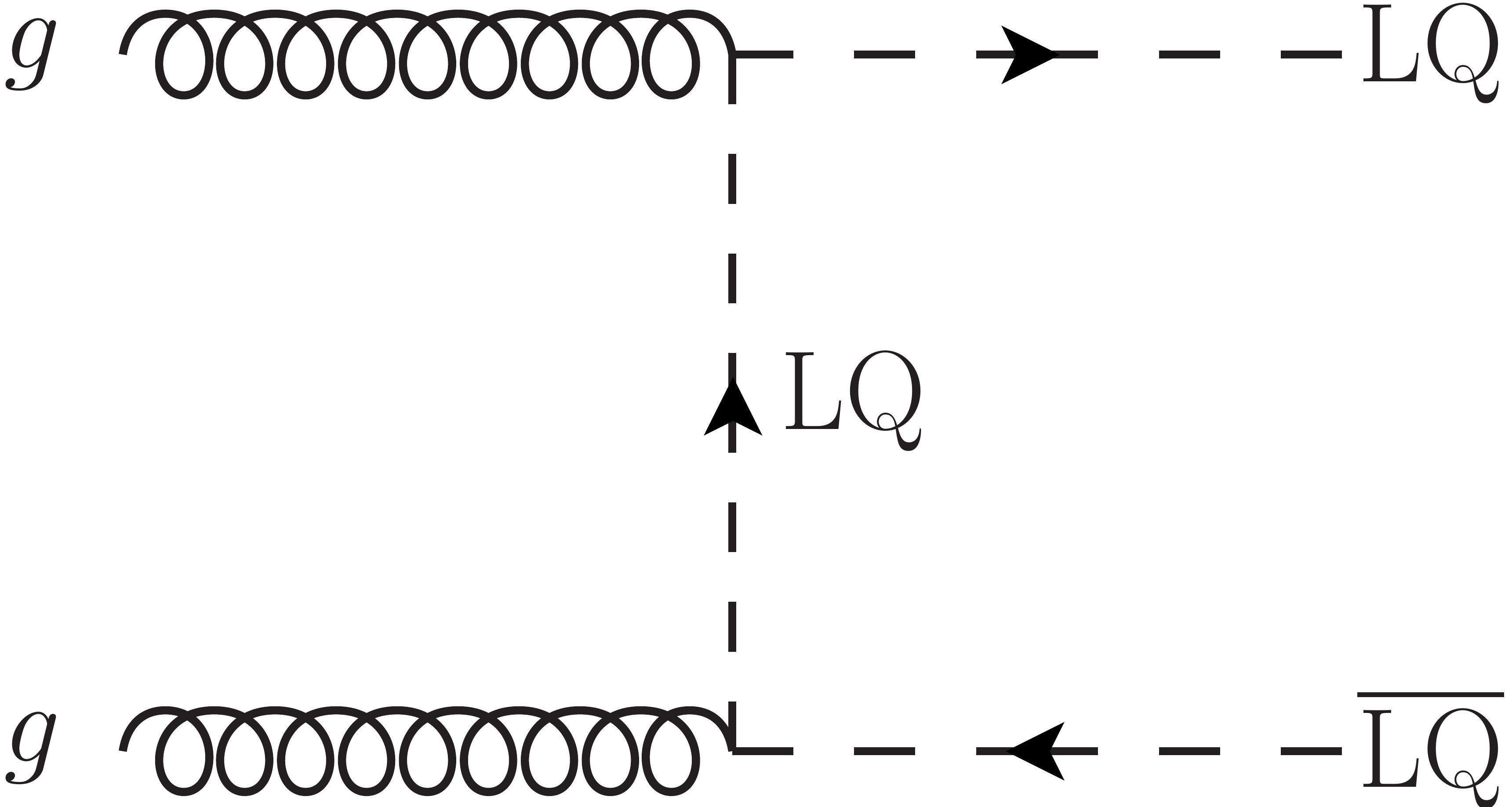}}\\
   \vspace{0.5cm}
   {\includegraphics[width=.3\textwidth]{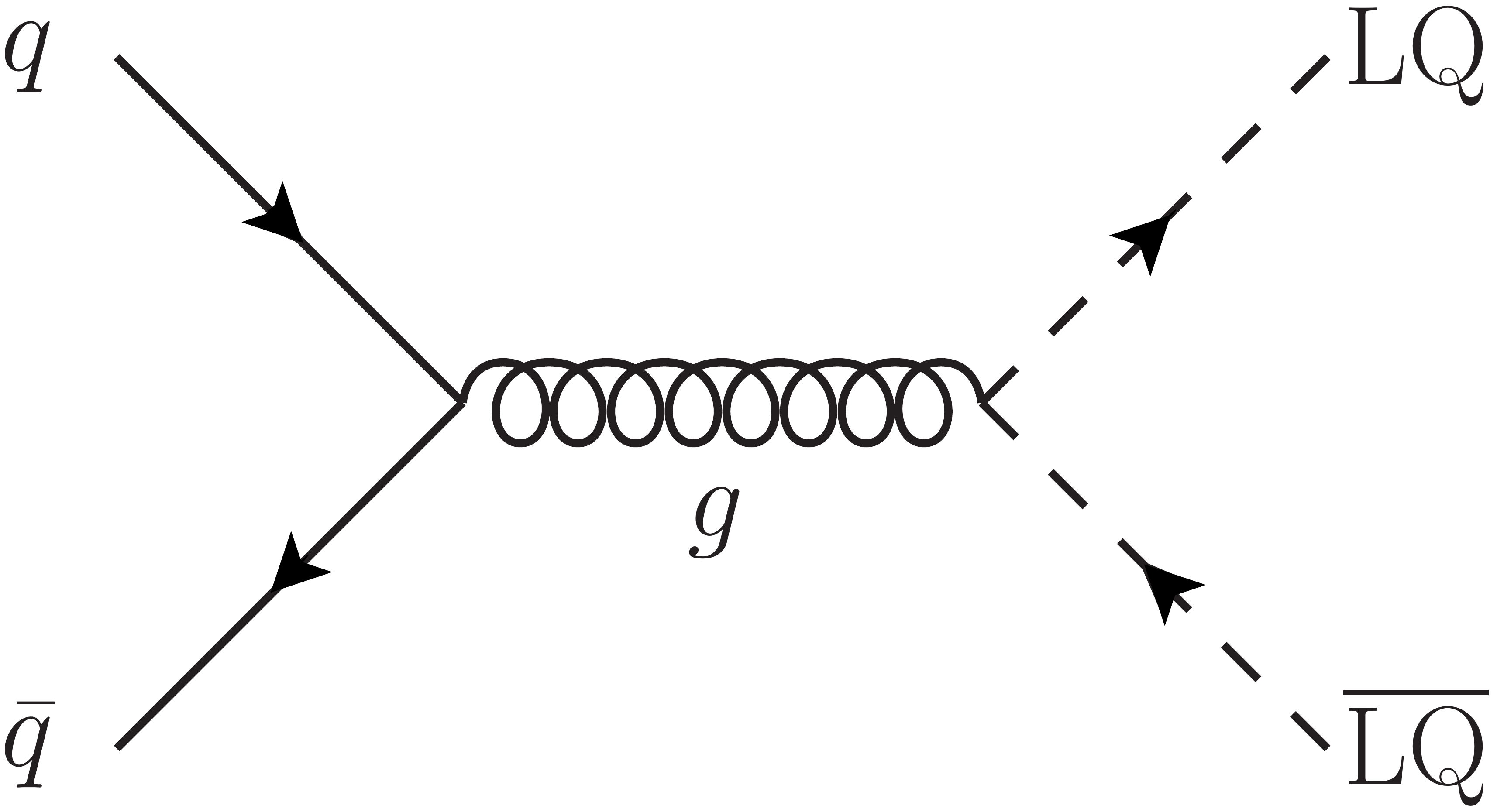}}
   \caption{Leading order Feynman diagrams for the scalar LQ pair production
     channels at the LHC.}
   \label{fig:lqDiagrams}
\end{figure*}

This analysis is based on data recorded in \Pp\Pp\ collisions
at $\sqrt{s} = 13\TeV$ with the CMS detector, corresponding
to an integrated luminosity of 35.9\fbinv. At LHC energies,
the pair production of LQs would mainly proceed via gluon-gluon fusion
with a smaller contribution from quark-antiquark annihilation. The
corresponding Feynman diagrams are shown in Fig.~\ref{fig:lqDiagrams}.
The production cross section as a function of \mlq\ has been calculated
at next-to-leading order (NLO) in perturbation theory~\cite{kramer}.
At the LHC, the LQ-lepton-quark Yukawa coupling has negligible
effect on the production rate for promptly decaying LQs, which are the
focus of our search.

The paper is organized as follows. Section~\ref{sec:cmsDetector} introduces
the CMS detector, and Sec.~\ref{sec:dataAndMC} describes the data and
simulated samples used in the search. The core of the analysis in terms of
event reconstruction and selection is discussed in Sec.~\ref{sec:eventSelection},
while the background estimation is presented in Sec.~\ref{sec:backgrounds}.
Section~\ref{sec:systematics} deals with the systematic uncertainties affecting
this analysis. Sections~\ref{sec:results} and \ref{sec:RPVSUSY} describe the
results of the LQ search and its interpretation in an exotic scenario of
supersymmetry, respectively. We conclude with a summary of the main results
in Sec.~\ref{sec:summary}.

\section{The CMS detector}
\label{sec:cmsDetector}

The key feature of the CMS apparatus is a superconducting solenoid of
6\unit{m} diameter, providing a magnetic field of 3.8\unit{T}. Within the
solenoid volume lie a silicon pixel and microstrip tracker, a lead-tungstate
crystal electromagnetic calorimeter (ECAL), and a brass-scintillator
hadron calorimeter (HCAL), each composed of a barrel and two end-cap sections.
Forward calorimeters extend the pseudorapidity ($\eta$) coverage provided
by the barrel and end-cap detectors. Muons are detected in gas-ionization
chambers embedded in the steel flux-return yoke outside the solenoid.
The first level of the trigger system~\cite{Khachatryan:2016bia}, composed of
custom electronics, uses information from the calorimeters and muon detectors
to select the most interesting events in an interval of less than 4\unit{$\mu$s}.
The high-level trigger processor farm further reduces the event rate from
around 100\unit{kHz} to 1\unit{kHz}, before data storage.  A detailed
description of the CMS detector, along with a definition of the coordinate
system used and the relevant kinematic variables, can be found in
Ref.~\cite{Chatrchyan:2008zzk}.

\section{Data and simulated samples}
\label{sec:dataAndMC}

Events are selected using a combination of triggers requiring either a single
electron or a single photon. Electron candidates are required to have a
minimum \pt of 27\,(115)\GeV for the low (high) threshold trigger.  Each
of these triggers examines clusters of energy deposited in the ECAL
that are matched to tracks reconstructed within a range $\abs{\eta}<2.5$.
Cluster shape requirements as well as calorimetric and track-based isolation
(only for the low threshold trigger) are also applied. By comparison,
the photon trigger requires $\pt>175\GeV$ without any requirements on
track-cluster matching, cluster shape, or isolation. The latter three
criteria are applied to electron triggers to reduce background rates
and are not necessary at high \pt. Therefore, the single photon and
electron triggers are combined to improve efficiency at high electron \pt.
Events selected using other single-photon triggers with lower thresholds
are used for determining the multijet background.

Monte Carlo (MC) simulation samples of scalar LQ signals are
generated using \PYTHIA version~8.212~\cite{pythia8p2} at leading
order (LO) with the \textsc{NNPDF2.3LO} parton distribution function
(PDF) set~\cite{nnpdf2p3lo}. Samples are generated for \mlq\ ranging
from 200 to 2000\GeV in 50\GeV steps.  The LQ is assumed to have
quantum numbers corresponding to the combination of an electron
($L=1$) and an up quark ($B=1/3$), implying it has an electric
charge of $-1/3$.
Possible formation of hadrons containing LQs is not included in the
simulation. The cross sections are normalized to the values
calculated at NLO~\cite{lqCrossSection,kramer} using the CTEQ6L1
PDF set~\cite{CTEQ6L1}.

The main backgrounds for searches in the \eejj\ and \enujj~channels
include Drell--Yan ($\cPZ/\PGg^{*}$) production with jets, top quark pair
production (\ttbar), single top quark and diboson ($\vv = \PW\PW$, \PW\cPZ, or
\cPZ\cPZ) production.  Additional background contributions arise from \wjets,
\gammajets, and multijet production, where jets are misidentified as electrons.
The \ttbar background in the \eejj~channel as well as the multijet background
in both channels are estimated from data, while MC simulated events are used
to calculate all other backgrounds. The \zjets, \wjets, and \vv samples are
generated at next to leading order (NLO) with \MGvATNLO version 2.3.3
using the FxFx merging method~\cite{Alwall:2014hca,amcAtNLOFXFX}.
Both \ttbar\ and single top quark events are generated at NLO using \MGvATNLO, and
\POWHEG v2 complemented with {\textsc{MadSpin}\xspace}~\cite{Artoisenet:2012st},
except for single top quark production in association with a \PW{} boson, where
events are generated with \POWHEG v1 at NLO~\cite{Nason:2004rx,Frixione:2007vw,Alioli:2010xd,Alioli:2009je,Re:2010bp,Frixione:2007nw},
and $s$-channel single top quark production, where \MGvATNLO at NLO is used.
The \gammajets\ events are generated with \MGvATNLO at LO with MLM
merging~\cite{Alwall:2007fs}. The NNPDF3.0 at NLO~\cite{nnpdf} PDF set
is used, except for \gammajets events that are generated using the LO PDF set.

The \wjets and \zjets samples are normalized to next-to-NLO (NNLO) inclusive cross
sections calculated with \FEWZ versions 3.1 and 3.1.b2, respectively~\cite{FEWZ}.
Single top quark samples are normalized to NLO inclusive cross sections~\cite{Aliev:2010zk,Kant:2014oha},
except for the $\PQt\PW$ production, where the NNLO calculations of Refs.~\cite{Kidonakis:2013zqa} are used.
The calculations from Refs.~\cite{tt1,tt2,tt3,tt4,tt5,tt6,tt7} with \textsc{Top++}2.0 are used
to normalize the \ttbar sample at NNLO in quantum chromodynamics (QCD) including resummation of the
next-to-next-to-leading-logarithmic soft gluon terms.

\PYTHIA\ 8.212 with the CUETP8M1 underlying event tune~\cite{cuetp8m1Tune} is used for hadronization and fragmentation in all
simulated samples, with the exception of a dedicated tune used for the \ttbar sample~\cite{cuetp8m2t4tune}.
All samples include an overlay of minimum bias events (pileup), generated
with an approximate distribution for the number of additional $\Pp\Pp$ interactions
expected within the same or nearby bunch crossings,
and reweighted to match the distribution observed in data.
In all cases, the \GEANTfour software v.10.00.p02~\cite{geant1,geant2} is used to
simulate the response of the CMS detector.

\section{Event reconstruction and selection}
\label{sec:eventSelection}

A particle-flow (PF) algorithm~\cite{Sirunyan:2017ulk} aims to reconstruct
and identify each individual particle in a given event, by optimally
combining information from the various elements of the CMS detector.
The energy of photons is directly obtained from the ECAL measurement.
On the other hand, the energy of
electrons is determined from a combination of their momentum at the primary
interaction vertex as determined by the tracker, the energy of the
corresponding ECAL clusters, and the energy sum of all bremsstrahlung
photons spatially compatible with originating from the associated track.
The momentum of muons is obtained from the curvature of the corresponding
track. The energy of charged hadrons is determined from a combination of
their momentum measured in the tracker and the matching ECAL and HCAL
energy deposits, corrected for zero suppression effects as well as for
the response function of the calorimeters to hadronic showers. Finally,
the energy of neutral hadrons is obtained from the corresponding corrected
ECAL and HCAL energy.

Electrons are identified by spatially matching a reconstructed
charged-particle track to a cluster of energy deposits in the ECAL.
The ECAL cluster is required to have longitudinal and transverse
profiles compatible with those expected from an electromagnetic shower.
Electrons used in this analysis are required to have
$\pt>50\GeV$ and $\abs{\eta}<2.5$,
excluding the transition regions between barrel and end-cap
detectors $1.4442<\abs{\eta}<1.5660$.
Additional selection criteria are applied to electron candidates in
order to reduce backgrounds while maintaining high efficiency for
identification of electrons with large \pt~\cite{Khachatryan:2015hwa}.
The absolute difference in $\eta$ between the ECAL cluster seed and the matched
track is required to be less than 0.004\,(0.006) in the barrel (end cap), and
the corresponding quantity in the azimuthal angle, $\phi$, must be less
than 0.06\unit{rad}.  Leptons resulting from the decay of LQs are expected
to be isolated from hadronic activity in the event.  Requirements
are, therefore, applied based on calorimeter energy deposits and tracks
in the vicinity of electron candidates.  The scalar sum of \pt
associated with calorimeter clusters in a cone of radius
$\Delta R=\sqrt{\smash[b]{(\Delta\eta)^2+(\Delta\phi)^2}}=0.3$
centered on the electron candidate, excluding clusters associated
to the candidate itself, must be less than 3\% of the
electron \pt.  A correction to the isolation sum accounts for
contributions from pileup interactions. The track-based isolation, calculated
as the scalar \pt sum of all tracks in the cone defined above, must
be less than 5\GeV to reduce misidentification of jets as electrons.
At most one layer of the pixel detector may have missing hits along
the trajectory of the matched track.  The track must also be compatible
with originating from the  primary \Pp\Pp\ interaction vertex,
which is taken to be the reconstructed vertex with the largest value
of summed physics-object $\pt^2$.
Here the physics objects are the jets, reconstructed using the
algorithm~\cite{Cacciari:2008gp,Cacciari:2011ma} with the tracks
assigned to the vertex as inputs, and the negative vector sum
of the \pt of those jets.
To correct for the possible difference of electron reconstruction
and identification efficiencies between collision and simulated data,
appropriate corrections or scale factors are applied to the simulated
samples.

Muons are used in defining a control region to estimate the \ttbar\
background contribution. They are identified as tracks in the central
tracker consistent with either a track or several hits in the muon
system~\cite{Sirunyan:2018fpa}. These muon candidates must have
$\pt>35\GeV$ and $\abs{\eta}<2.4$, and are required to pass a series
of identification criteria designed for high-\pt muons as follows.
Segments in at least two muon stations must be geometrically matched
to a track in the central tracker, with at least one hit from a muon chamber
included in the muon track fit.
In order to reject muons from decays in flight and increase momentum
measurement precision, at least five tracker layers must have hits
associated with the muon, and there must be at least one hit in the pixel detector.
Isolation is imposed by requiring the \pt sum of tracks in a cone of
$\Delta R = 0.3$ (excluding the muon itself) divided by the muon \pt to be less
than 0.1. For rejection of cosmic ray muons, the transverse impact parameter
of the muon track with respect to the primary vertex must be less than
2~mm and the longitudinal distance of the track formed from tracker
system only to the primary vertex must be less than 5\unit{mm}.
Finally, the relative uncertainty on the \pt measurement from the muon track must be less than 30\%.

Jets are reconstructed using the anti-\kt
algorithm~\cite{Cacciari:2008gp,Cacciari:2011ma} with a distance parameter
of 0.4.  Their momentum is determined as the vectorial sum of all particle
momenta in the jet, and is found in simulation to be within 5--10\% of the
true momentum~\cite{CMS-PAS-JME-16-003} over the entire \pt spectrum and
detector acceptance. Pileup interactions can contribute spurious tracks and
calorimeter energy deposits to the jet momentum. To mitigate this effect,
tracks identified to be originating from pileup vertices are discarded,
while a correction~\cite{Khachatryan:2016kdb} is applied
to compensate for the remaining contributions. Jet energy corrections
are extracted from simulation to compensate for differences between
the true and reconstructed momenta of jets. In situ measurements of the
momentum balance in dijet, \gammajets, \zjets, and multijet events are
used to estimate and correct for any residual differences in jet energy
scale between data and simulation~\cite{Khachatryan:2016kdb}. Additional
selection criteria are applied to all jets to remove those potentially
affected by spurious energy deposits originating from instrumental
effects or reconstruction failures~\cite{met_performance}. Jets must
have $\pt>50\GeV$ and $\abs{\eta}<2.4$, and only jets separated
from electrons or muons by $\Delta R>0.3$ are retained.

The missing transverse momentum (\ptvecmiss) is given by the negative vector
sum of \pt of all PF candidates in the event. The magnitude of \ptvecmiss is
referred to as \ptmiss.

To identify $\cPqb$ jets arising from top quark decays
in the determination of the \enujj~background control regions, the combined
secondary vertex algorithm is used with the loose working point of
Ref.~\cite{Sirunyan:2017ezt}. Based on simulation, the corresponding
$\cPqb$-jet identification efficiency is above 80\% with a probability
of 10\% of misidentifying a light-flavor jet.

\subsection{The \texorpdfstring{\eejj}{eejj}~channel}
\label{subsec:eventSelEejj}

For the \eejj~analysis, we select events with at least two
electrons and at least two jets passing the criteria described
above. No charge requirements are imposed on the electrons.
When additional objects satisfy these requirements, the
two highest \pt electrons and jets are considered.
Further, there should not be any muon fulfilling the requirements
mentioned earlier in this section.
The dielectron invariant mass \mee is required to be greater than 50\GeV.
The \pt of the dielectron system must be greater than 70\GeV.
The scalar \pt sum over the electrons and two jets,
$\st = \pt(\Pe_1)+\pt(\Pe_2)+\pt(\text j_1)+\pt(\text j_2)$, must be at
least 300\GeV. This initial selection is used for the determination of
backgrounds in control regions, as explained in Section~\ref{sec:backgrounds}.

\begin{figure}[htb]
  \centering
    {\includegraphics[width=.49\textwidth]{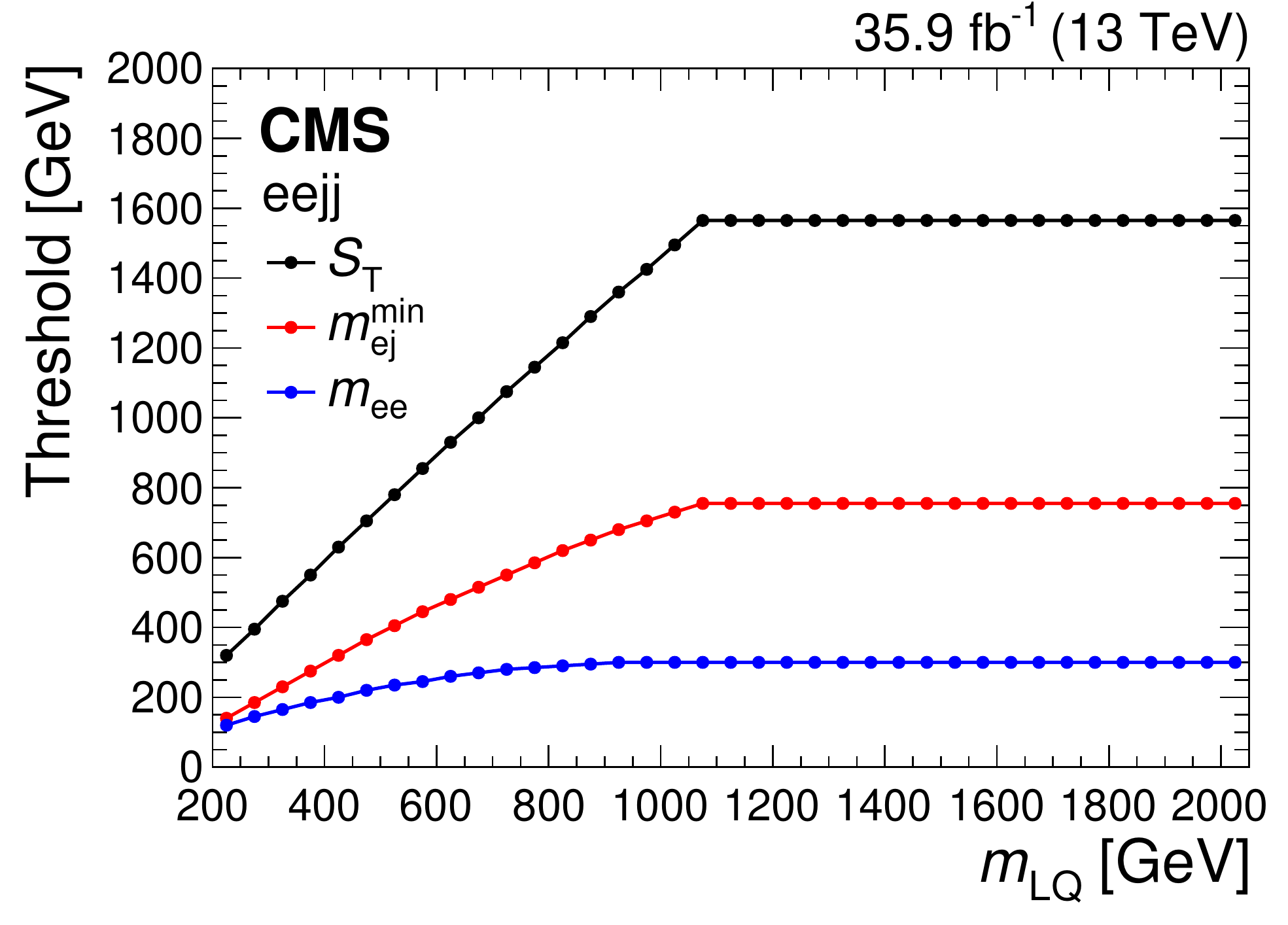}}
    {\includegraphics[width=.49\textwidth]{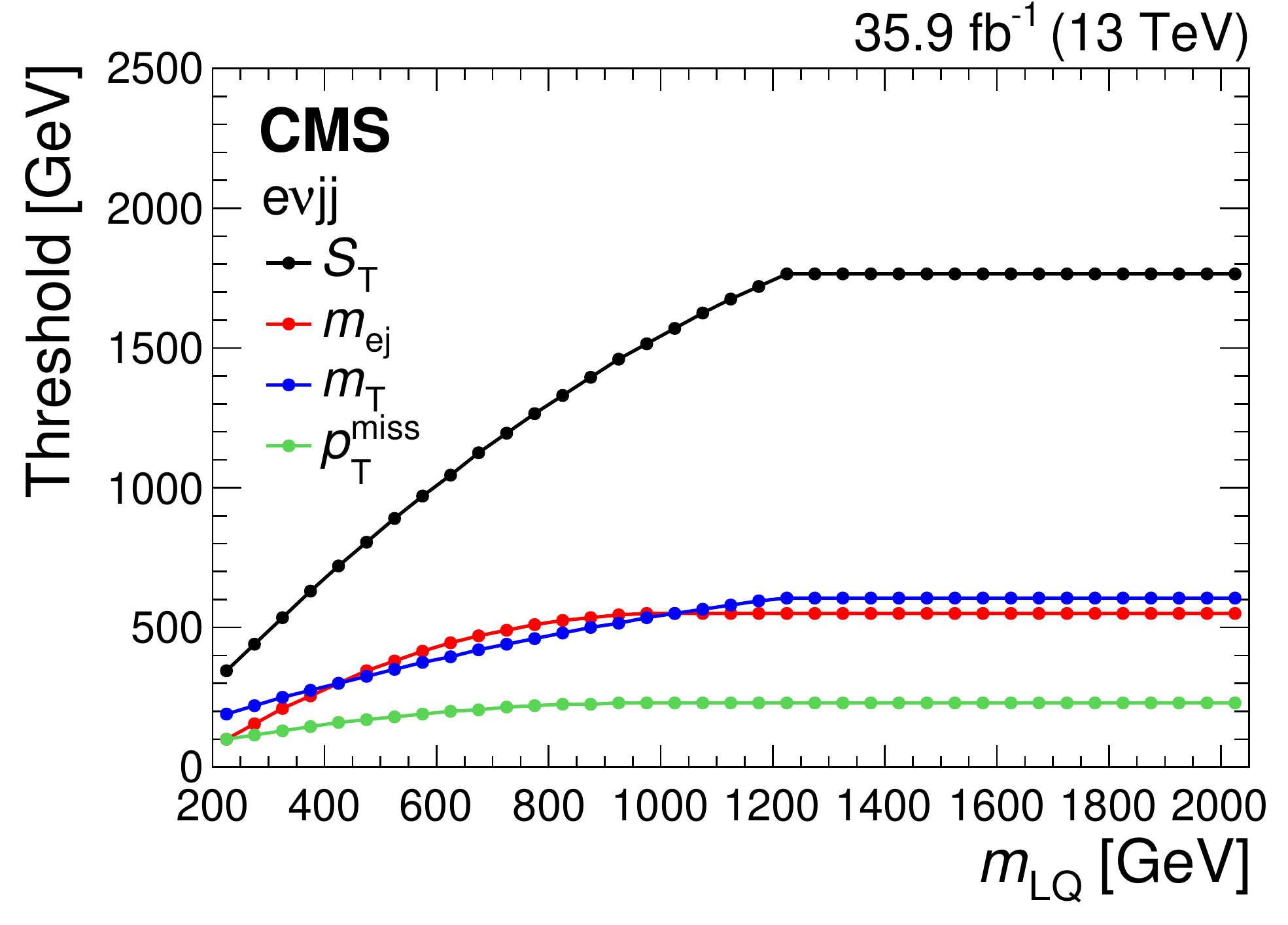}}
    \caption{
      Optimized threshold values applied for the selection variables in the
      \eejj\ (\cmsLeft) and \enujj\ (\cmsRight) channels as a function of \mlq.
    }
    \label{fig:optimizationThresholds}
\end{figure}

Final selections are then optimized by
maximizing the Punzi criterion for observation of a signal at a
significance of five standard deviations~\cite{punzi}.  These
selections are determined by examining three variables: \mee, \st,
and \mejmin.  The electron-jet pairing is chosen to minimize
the difference in the invariant mass of the LQ candidates, and
the quantity \mejmin is defined as the smaller of the
two masses.  Thresholds for the three observables are varied independently, and
the Punzi criterion is then calculated at each set of thresholds
as well as for each \mlq\ hypothesis. The optimized thresholds
as a function of \mlq\ are shown in Fig.~\ref{fig:optimizationThresholds}
(\cmsLeft). For the \mlq\ hypotheses above 1050\GeV, the statistical
uncertainty in the background prediction becomes large, making an
optimization for these masses impossible, and thus the thresholds
for the 1050\GeV hypothesis are applied.

\subsection{The \texorpdfstring{\enujj}{enujj}~channel}
\label{subsec:eventSelEnujj}

In the \enujj~channel, we select events containing exactly one
electron, at least two jets, and $\ptmiss>100\GeV$.
The electron and jets must pass the aforementioned identification criteria.
Events with isolated muons are rejected, applying the same
criteria as for the \eejj\ channel.
The absolute difference in the angle between the \ptvecmiss
and the leading \pt jet, $\Delta\phi(\ptvecmiss,\text j_1)$, is required
to be larger than 0.5\unit{rad}.  This helps reject events with \ptmiss
arising primarily from instrumental effects.
The $\Delta\phi(\ptvecmiss,\Pe)$ must be greater than 0.8\unit{rad} for
similar reasons.
The \pt and transverse mass of the \ptvecmiss-electron system must be greater
than 70 and 50\GeV, respectively. Here and later, the transverse mass of a
two-object system is given by $\mT=\sqrt{\smash[b]{2p_{{\rm T},1}p_{{\rm T},2}
(1-\cos\Delta\phi)}}$, with $\Delta\phi$ being the angle between the \pt
vectors of two objects, namely \ptvecmiss, electron and jet. The \mT\ criterion
helps suppress the \wjets\ contribution. Finally, selected events must have
$\st>300\GeV$, where $\st = \pt(\Pe)+\ptmiss+\pt(\text j_1)+\pt(\text j_2)$.
This initial selection is used for the determination of backgrounds in control regions,
similarly to the \eejj\ channel.

The selection criteria are then optimized in a similar fashion as for
the \eejj\ channel, except that four observables are considered for
final selections at each \mlq\ hypothesis: \st, \mT\ of the \ptvecmiss-electron
system, \ptmiss, and the electron-jet invariant mass \mej.
The \ptvecmiss-jet and electron-jet pairing is chosen to minimize the
difference in \mT\ between the two LQ candidates. The optimized thresholds as a
function of \mlq\ are shown in Fig.~\ref{fig:optimizationThresholds} (\cmsRight).
As with the \eejj\ channel, for the \mlq\ hypotheses above 1200\GeV, the
thresholds for the 1200\GeV hypothesis are used.

\section{Background estimation}
\label{sec:backgrounds}

The SM processes that produce electrons and jets can have final
states similar to those of an LQ signal and are, therefore, considered
as backgrounds for this search.  These include dilepton events
from \zjets, \ttbar, and \vv; single top quark production; and \wjets.
Another background arises from multijet production in which at least one
jet is misidentified as an electron.

The major backgrounds in the \eejj channel are \zjets and
\ttbar production.  The \zjets\ background is estimated from simulation
and normalized to the data in a control region that comprises the
initial selection plus a window of $80<\mee<100\GeV$ around the
nominal \PZ\ boson mass; the latter criterion is applied to enrich
the sample with \zjets\ events.
The \mee\ distribution is corrected for the presence of
non-\zjets\ events in the data control region using simulation.
The resulting normalization factor applied to the \zjets\ simulated events
is $R_{\PZ} = 0.97 \pm 0.01\stat$.

The contribution from \ttbar\ events containing two electrons is estimated
using a control region in data, which consists of events containing one
electron and one muon, to which all applicable \eejj selection criteria
are applied. Residual backgrounds from other processes are subtracted using
simulated event samples. Corrections for the branching fractions between
the two states as well as for the differences in electron/muon identification
and isolation efficiencies and acceptances are determined using simulation.
The difference in the trigger efficiency between the one- and
two-electron final states is corrected by reweighting each event in
the \Pe\Pgm\ sample with the calculated efficiencies for the single
electron final state.

After application of event selection requirements, the
background contribution to the \eejj channel arising from single top
quark production, \wjets, and \vv is found to be small and is estimated
from simulations.

The multijet background in the \eejj\ channel is estimated using control
samples in data.
The electron identification requirements for
the calorimeter shower profile and track-cluster matching are relaxed
to define a loose selection.
We measure the probability that an electron candidate that passes
the loose selection requirements also satisfies the electron
identification and isolation criteria used in the analysis. This
probability is obtained as a function of the candidate \pt\ and
$\eta$. The events are required to have exactly one loose
electron, at least two jets, and low \ptmiss ($<100\gev$).
Contributions from electrons satisfying the full identification
requirements are removed. The number of such electrons is calculated
by comparing the number of candidates that pass the tight selection
criteria minus the track-isolation requirement, with those that
satisfy the track-isolation requirement but fail one of the other
selection criteria.
This sample is dominated by QCD multijet events. The distribution
of multijet events in the \eejj\ channel following final selections
is obtained by applying the measured probability twice to an event sample
with two electrons passing loose electron
requirements, and two or more jets that satisfy all the requirements
of the signal selection.  The normalization is obtained by scaling the
weighted multijet sample to an orthogonal control region defined by
inverting track-isolation requirement for electrons.

\begin{figure}[hbtp]
  \centering
    {\includegraphics[width=.49\textwidth]{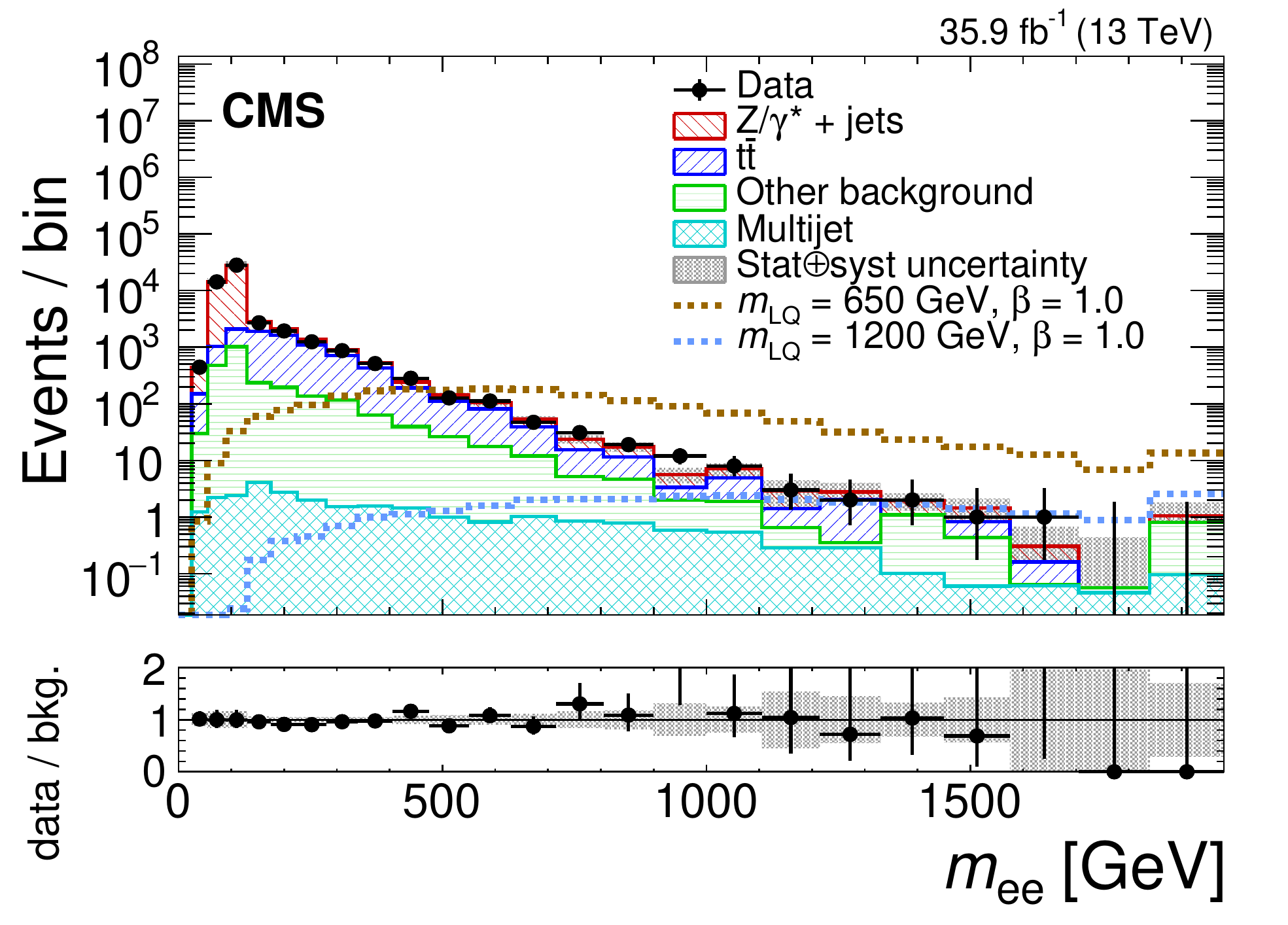}}\\
    {\includegraphics[width=.49\textwidth]{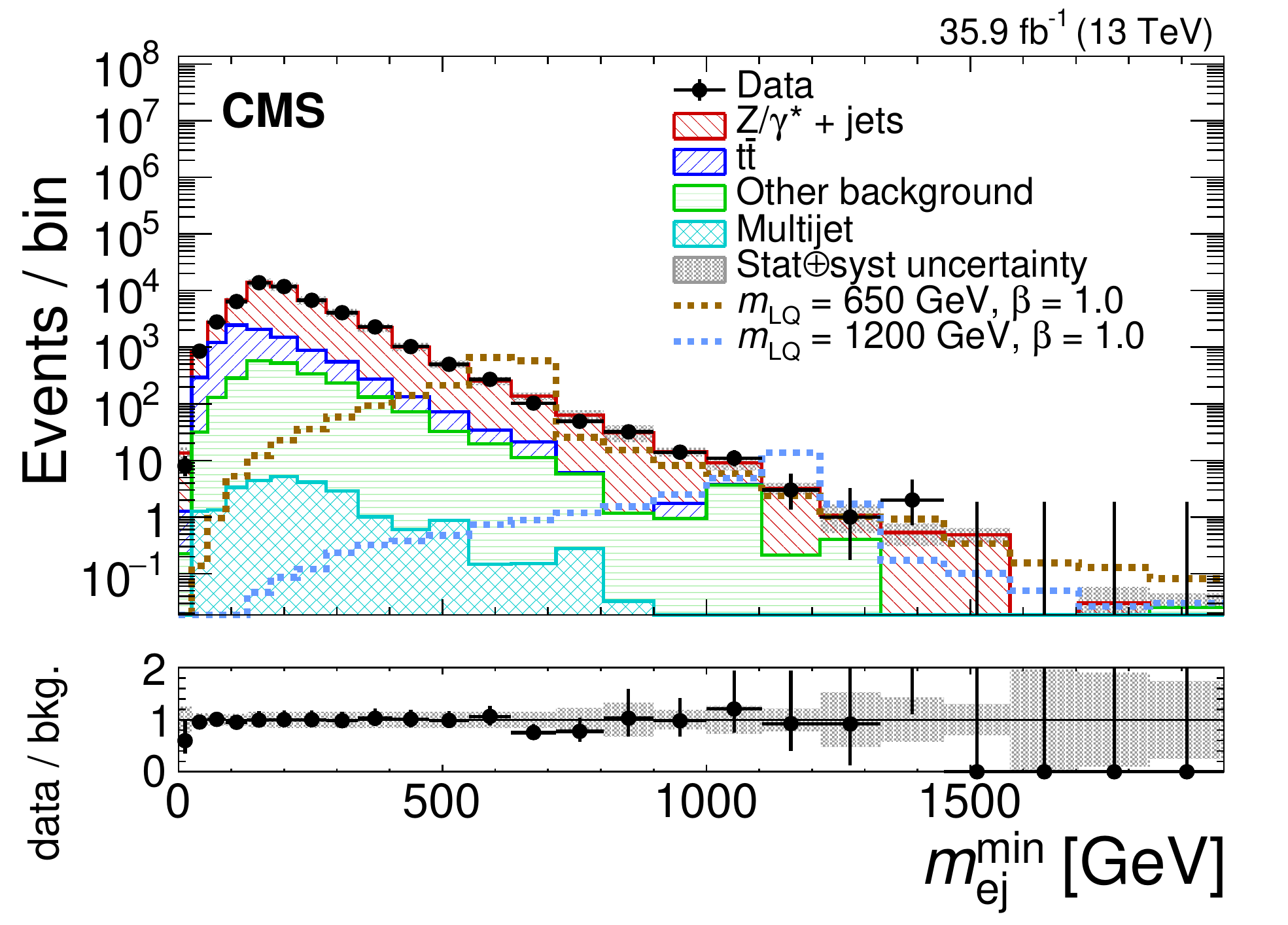}}
    {\includegraphics[width=.49\textwidth]{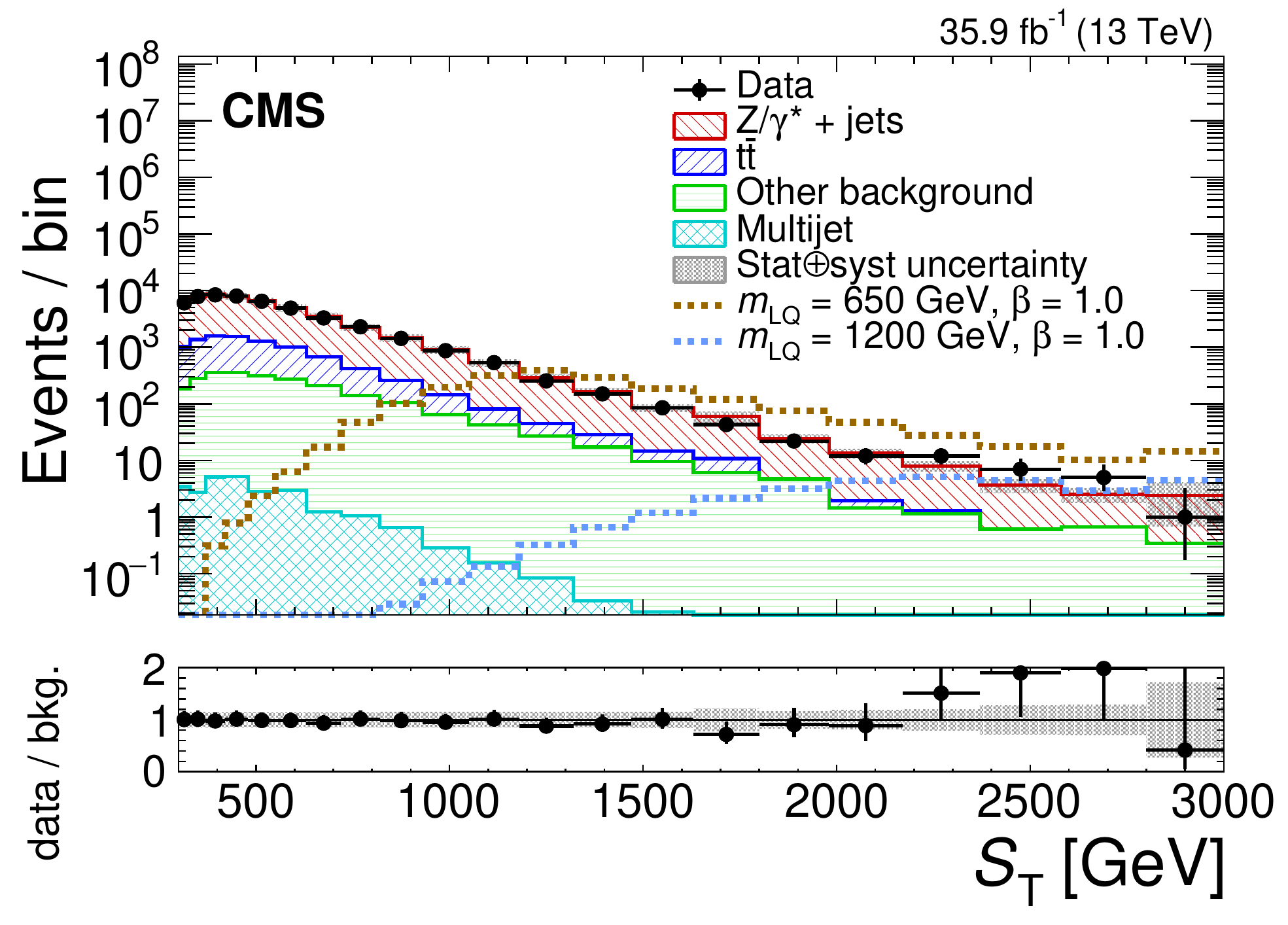}}
    \caption{Data and background comparison for events passing the
      initial selection requirements for the \eejj~channel,
      shown for the variables used for final selection optimization:
      \mee~(upper), \mejmin~(\cmsMiddle), and \st~(\cmsRightThree).
      ``Other background" includes diboson, single top quark, and \wjets.
      Signal predictions for $\mlq= 650$ and 1200\GeV hypotheses are
      overlaid on the plots. The last bin includes all events
      beyond the upper $x$-axis boundary.
    }
    \label{fig:eejjPreselPlots}
\end{figure}

Distributions of kinematic variables for the \eejj channel in data,
including those used in the final selections, have been studied
at the initial selection level, and are found to agree with the
background models within background estimation uncertainties.
The distributions of \st, \mejmin, and \mee are shown
in Fig.~\ref{fig:eejjPreselPlots}.

The largest background in the \enujj channel comes from
\wjets\ and \ttbar\ production.  Single top quark, \vv, and \zjets backgrounds
have small contributions and are estimated from simulations.
The QCD multijet background is estimated from
control samples in data using the same probability for jets to be
misidentified as electrons as is used
in the background estimation for the \eejj channel.
The number of multijet events at the final selection is obtained by
selecting events having exactly one loose electron, large \ptmiss,
and at least two jets satisfying the signal selection criteria, and
weighting these with the probability of a jet being misidentified
as an electron.

The background contributions from \wjets and \ttbar are estimated from
simulation, and normalized to the data in a control region defined
by requiring $50<\mT<110\GeV$ after the initial selection.
Then \cPqb-tagging information is used to distinguish \wjets from \ttbar
in the control region.  The \wjets contribution is
enhanced by requiring zero \cPqb-tagged jets in the event, while the
\ttbar control region is defined by requiring at least one \cPqb-tagged
jet in the event. These regions each have a purity of about 75\%.
The normalization factors for the two backgrounds are
determined from these control regions using
\begin{linenomath}
  \begin{equation}
    \begin{aligned}
    & N_1 = R_{\ttbar} N_{1,\ttbar} + R_{\PW} N_{1,{\PW}} + N_{1,\mathrm{O}}\\
    & N_2 = R_{\ttbar} N_{2,\ttbar} + R_{\PW} N_{2,{\PW}} + N_{2,\mathrm{O}},
    \end{aligned}
    \label{bkeqn}
  \end{equation}
\end{linenomath}
where $N_{1\,(2)}$ is the number of events in the \ttbar\,(\wjets) control
region in data.
The terms $N_{i,\ttbar}$ and $N_{i,{\PW}}$ are the numbers of
\ttbar\ and \wjets\ events in the simulated samples, while $N_{i,\mathrm{O}}$
is the number of events arising from other background sources, namely
diboson, single top quark, \zjets\ and multijet. The subscript
$i=1,2$ refers to the two control regions described above.
The background normalization factors
$R_{\ttbar} = 0.92\pm0.01\stat$
and
$R_{\PW} = 0.87\pm0.01\stat$
are then determined by solving Eq.~(\ref{bkeqn}).

\begin{figure*}[htbp!]
  \centering
    {\includegraphics[width=.49\textwidth]{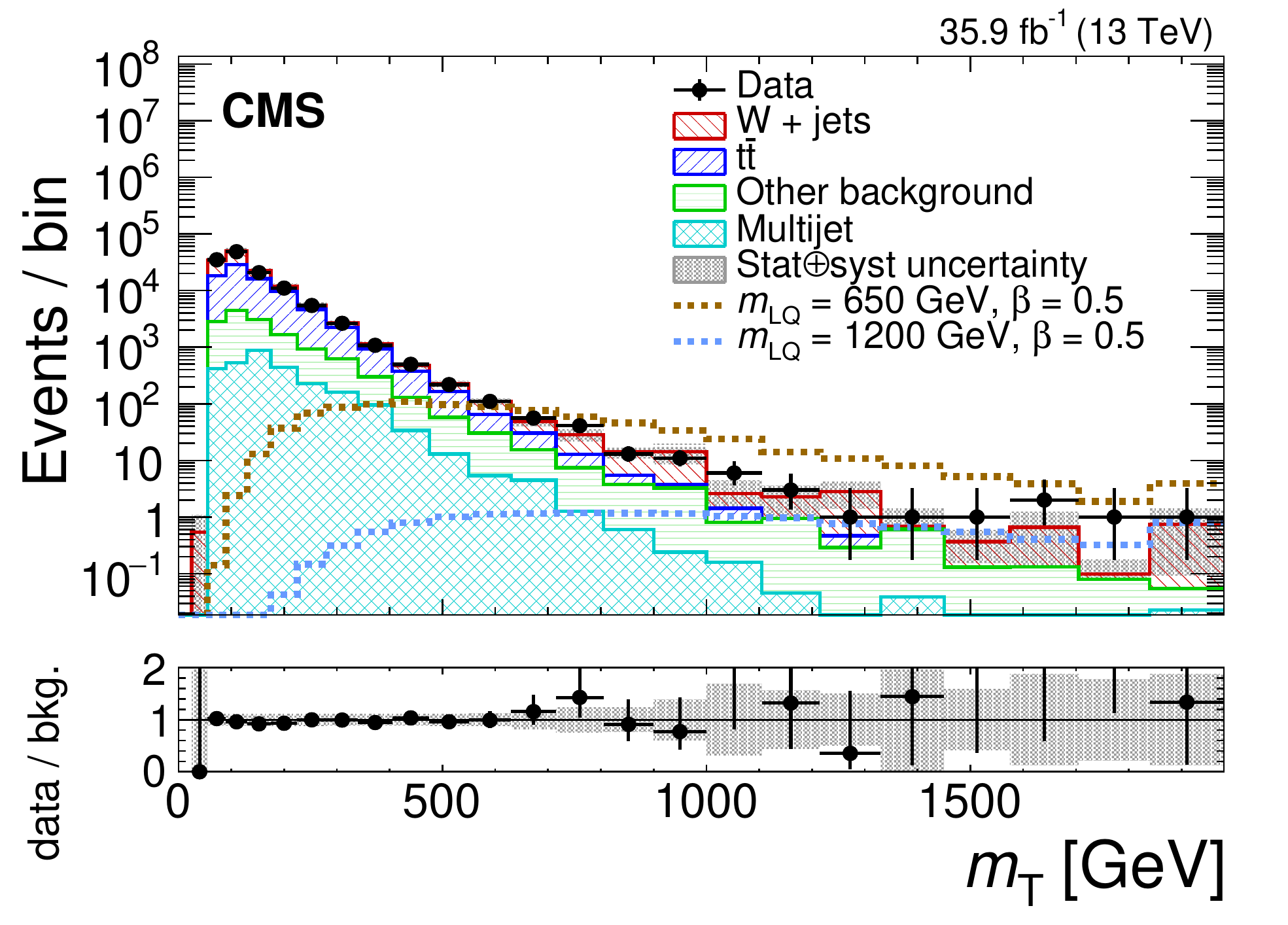}}
    {\includegraphics[width=.49\textwidth]{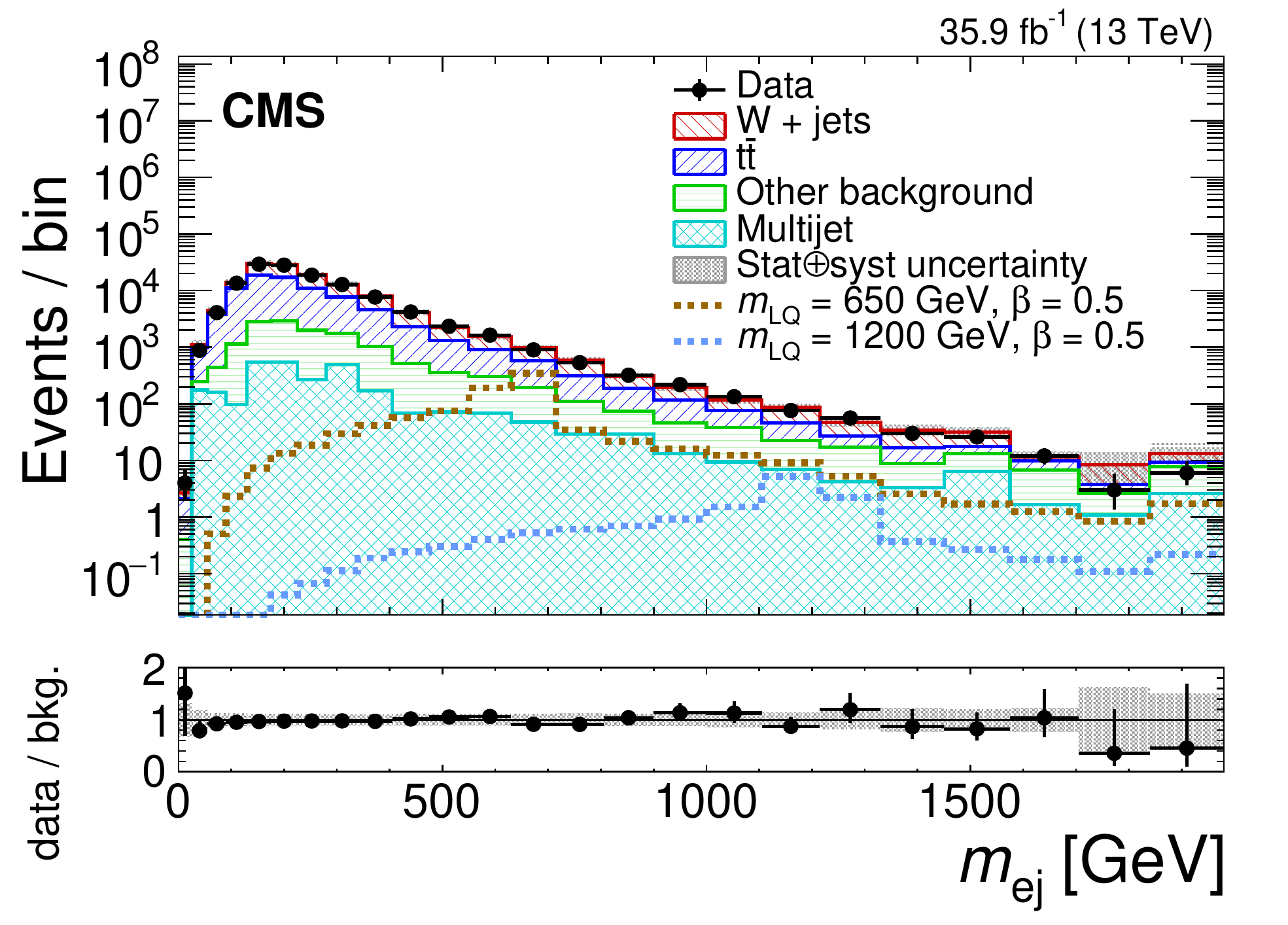}}
    {\includegraphics[width=.49\textwidth]{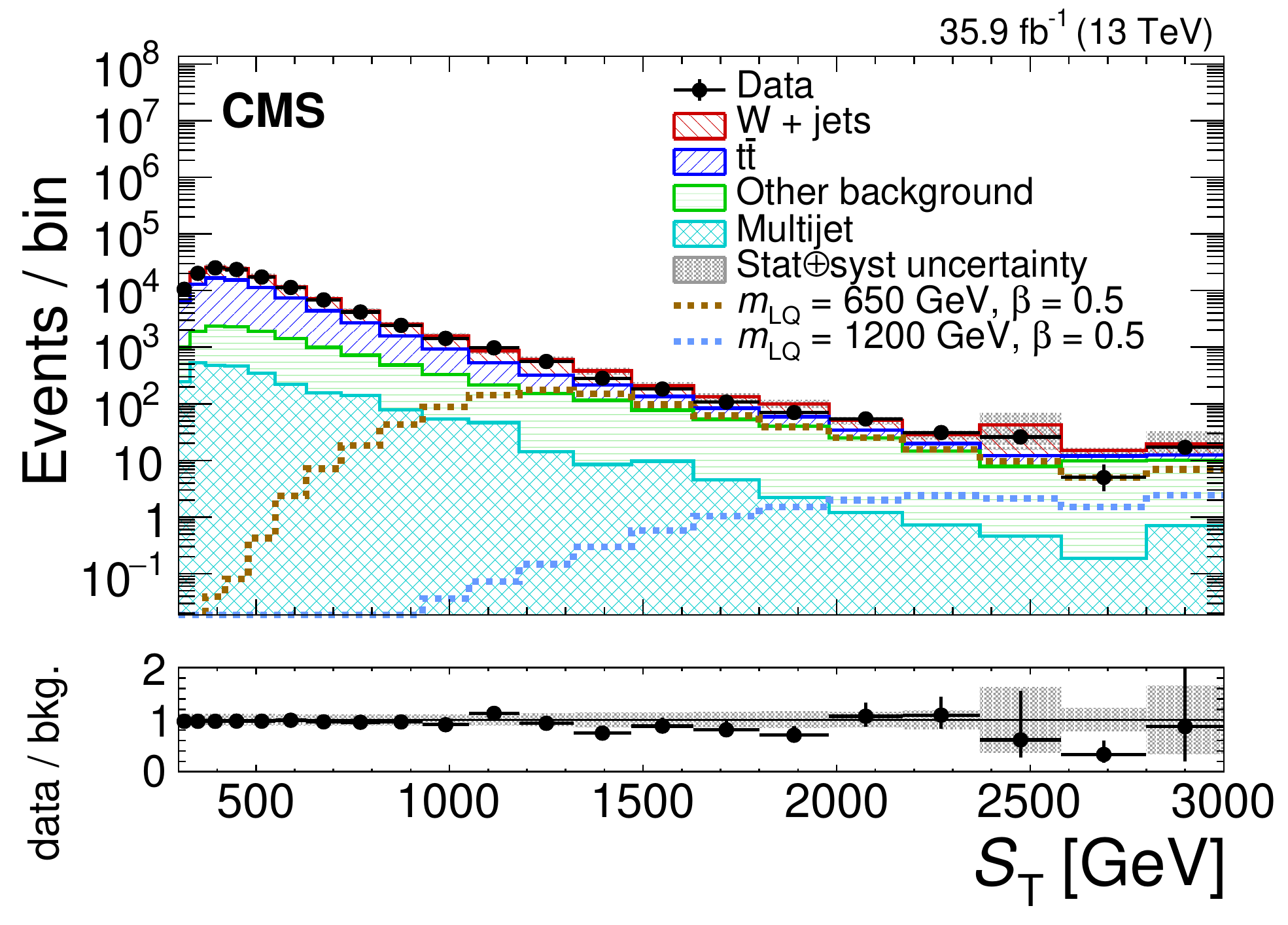}}
    {\includegraphics[width=.49\textwidth]{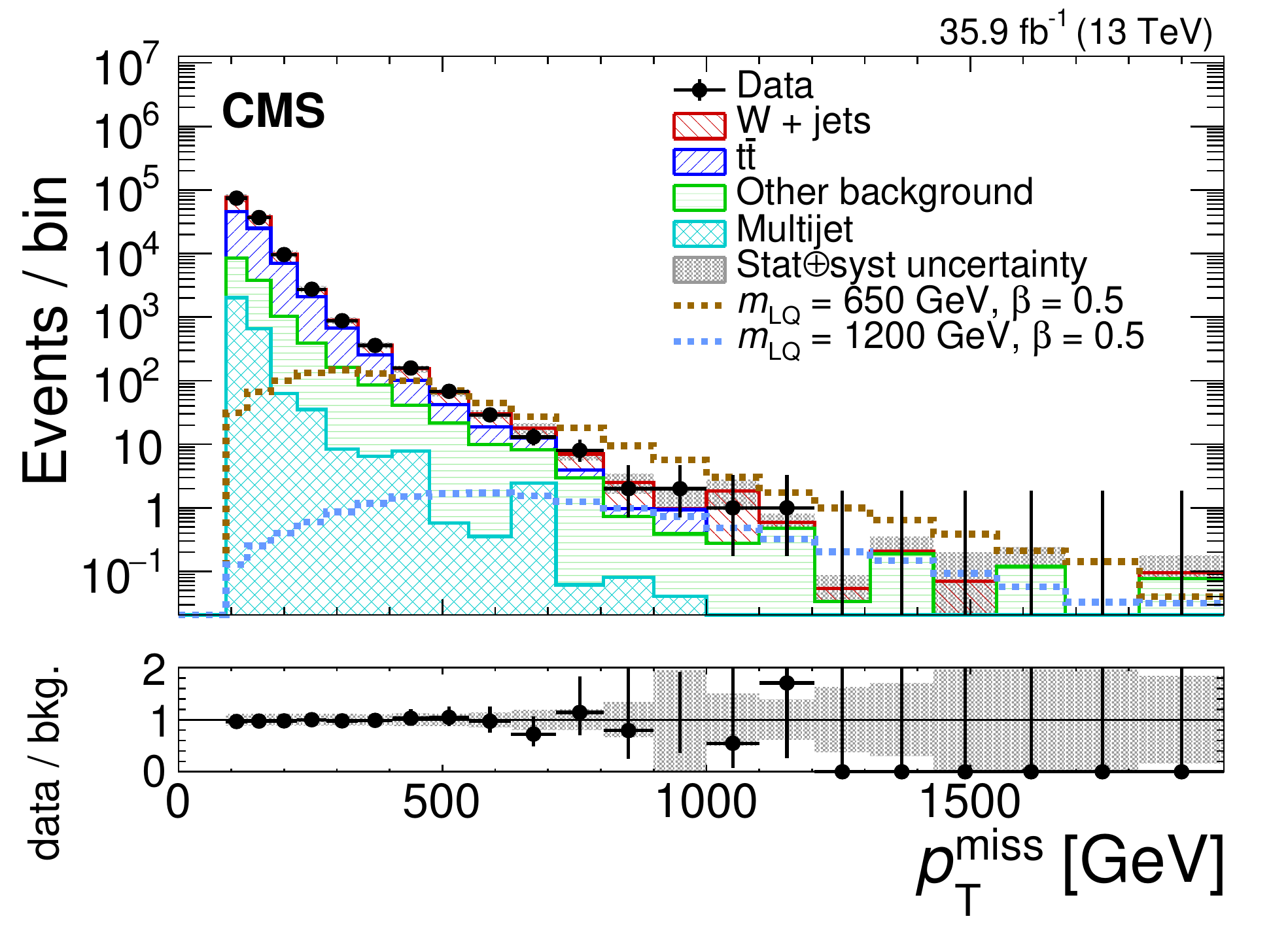}}
    \caption{Data and background for events passing the initial
      selection requirements in the \enujj~channel, shown for the
      variables used for final selection optimization: \mT~(upper left),
      \mej~(upper right), \st~(lower left), and \ptmiss~(lower right).
      ``Other background" includes diboson, single top quark, and \zjets.
      Signal predictions for $\mlq= 650$ and 1200\GeV hypotheses are
      overlaid on the plots. The last bin includes all events beyond
      the upper $x$-axis boundary.
    }
    \label{fig:enujjPreselPlots}
\end{figure*}

The observed distributions of kinematic variables for the \enujj channel following
the initial selection
are found to agree with the background prediction within estimation uncertainties.
The distributions of \mT, \mej, \st, and \ptmiss are
shown in Fig.~\ref{fig:enujjPreselPlots}.

\section{Systematic uncertainties}
\label{sec:systematics}

The sources of systematic uncertainties considered in this analysis
are listed in Table~\ref{tab:systematicLQ650}.   Uncertainties in the
reconstruction of electrons, jets and \ptmiss affect the selected sample of events
used in the analysis.
The uncertainty due to the electron energy scale is obtained by
shifting the electron energy up and down by 2\%. The uncertainty in the electron
energy resolution is measured by smearing the electron energy by $\pm$10\%~\cite{CMS:EGM-14-001}.
The uncertainties due to electron
reconstruction and identification efficiencies are obtained by
varying the corresponding scale factors applied to simulated
events by $\pm 1$ standard deviation with respect to their nominal values.
The trigger efficiency for electrons is measured by utilizing
the tag-and-probe method~\cite{Khachatryan:2010xn} in data, and parametrized
as a function of electron \pt and $\eta$.  The corresponding uncertainty depends on the
number of data events and is almost entirely statistical in origin for the
kinematic range studied in this analysis.

The uncertainty due to the jet energy scale is obtained
by varying the nominal scale correction by $\pm$1 standard
deviation and taking the maximum difference with
respect to the nominal event yield.  The jet energy resolution models the
variation between the reconstructed and generated jets. The corresponding
uncertainty is obtained by modifying the parametrization of this
difference~\cite{Khachatryan:2016kdb}.

To determine uncertainties in \ptmiss, we consider up and down shifts in
the jet energy scale and resolution, electron energy correction, and the
scale corrections applied to the energy not associated with any PF candidates.
For each variation, a new \ptmiss vector is computed for each event.
The uncertainties corresponding to different variations in the
quantities are then added in quadrature to determine the variation in \ptmiss,
and the maximum difference of the event yield with respect to
nominal is taken as the uncertainty.

Variations in the shape of the \zjets (\eejj~channel only), \wjets and
\ttbar (\enujj~channel only), and diboson (both channels) backgrounds are
determined using simulated samples with renormalization and factorization
scales independently varied up and down in the matrix element by a factor of two,
yielding eight different combinations.  The event yields are then
calculated for each of these variations and the maximum variation
with respect to nominal is taken as the systematic uncertainty.
The corresponding normalization uncertainties are
evaluated from the statistical uncertainties in the scale factors
obtained while normalizing these backgrounds to data in the control regions.
In the \enujj~channel, an additional uncertainty of 10\% is included
to account for the observed differences associated with the choice of
the \mT range, defining the control region used to calculate the normalization scale
factors. As described above, \cPqb-tagging is used to define the control region for
\wjets and \ttbar normalization in the \enujj channel; therefore the uncertainty
in the \cPqb-tagging efficiency (3\%) is taken into account.

The uncertainty in the QCD multijet background is assessed by using an
independent data sample. This sample is required to have exactly
two electron candidates satisfying loosened criteria applied to the
track-cluster matching, the isolation (both track-based and calorimetric),
and the shower profile.
We compare the number of events in this sample, where one candidate
satisfies the electron selection requirements, to that predicted by
the multijet background method.  This test is repeated on a subsample of
the data after applying an \st threshold of 320\GeV, which corresponds
to the optimized final selection for an LQ mass of 200\GeV. The relative
difference of 25\% observed between the results of the two tests
is taken as the systematic uncertainty in the probability for a jet
to be misidentified as an electron and applied in the \enujj channel.
For the \eejj case, we assume full correlation between the two
electrons and take 50\% as the uncertainty.

The uncertainty in the integrated luminosity is 2.5\%~\cite{CMSlumiPas}.
An uncertainty in the modeling of pileup is evaluated by reweighting the
simulated events after varying the inelastic $\Pp\Pp$
cross section by $\pm$4.6\%~\cite{Sirunyan:2018nqx}.
The acceptance for both signal and backgrounds, and the expected
background cross sections are affected by PDF uncertainties.
We estimate this effect by evaluating the
complete set of NNPDF~3.0 PDF eigenvectors, following the
PDF4LHC prescription~\cite{Botje:2011sn,Alekhin:2011sk,Whalley:2005nh,Bourilkov:2006cj,nnpdf}.

\begin{table*}[htb!]
  \topcaption{Systematic uncertainties for the \eejj~and \enujj~channels.
    The values shown are calculated for the selections used in the
    $\mlq=1000\GeV$ search hypothesis and reflect the variations in
    the event yields due to each source.
    Major backgrounds, namely \zjets\ (\eejj), \wjets and \ttbar
    (\enujj), are normalized at the initial selection level when
    calculating the effect of shifts for various systematics.
  }
  \centering
    \begin{scotch}{lrrrrr}
 {}          & \multicolumn{2}{c}{\eejj}         &  \multicolumn{2}{c}{\enujj} \\
Source of the uncertainty	&  Signal (\%)	&Background (\%)&  Signal (\%)	&Background (\%)\\ \hline
  Electron energy scale     	&	1.5	&	2.5	&	1.9	&	6.9	\\
  Electron energy resolution	&	0.2	&	5.3	&	0.1	&	4.9	\\
  Electron reconstr. efficiency &       3.0     &       3.0     &       0.6     &       0.8     \\
  Electron identif. efficiency  &       1.3     &       0.3     &       0.6     &       0.1     \\
  Trigger                   	&	1.1	&	1.4	&	9.5	&	7.6	\\
  Jet energy scale          	&	0.5	&	0.9	&	0.5	&	2.3	\\
  Jet energy resolution     	&	0.1	&	1.7	&	0.1	&	2.4	\\
  \ptmiss                       &       \NA     &       \NA     &       0.8     &       13.1    \\
  \zjets\ shape         	&	\NA	&	5.6	&	\NA	&	\NA	\\
  \zjets\ normalization         &       \NA     &       1.0     &       \NA     &       \NA     \\
  \wjets\ shape         	&	\NA	&	\NA	&	\NA	&	7.1	\\
  \wjets\ normalization         &       \NA     &       \NA     &       \NA     &       1.1     \\
  \wjets\ sideband selection	&	\NA	&	\NA	&	\NA	&	10.0	\\
  \wjets\ \cPqb~tagging    	&	\NA	&	\NA	&	\NA	&	3.0	\\
  \ttbar\ shape                 &       \NA     &       \NA     &       \NA     &       10.4    \\
  \ttbar\ normalization      	&	\NA	&	1.0	&	\NA	&	1.0	\\
  \ttbar\ \cPqb~tagging      	&	\NA	&	\NA	&	\NA	&	3.0	\\
  Diboson shape                 &       \NA     &       3.4     &       \NA     &       3.2     \\
  QCD multijet      	        &       \NA     &     $<$0.1    &       \NA     &       2.6     \\
  Integrated luminosity         &       2.5     &       0.6     &       2.5     &       0.5     \\
  Pileup                        &       0.2     &       1.0     &       0.4     &       1.4     \\
  PDF                           &       2.8     &       3.0     &       2.9     &       4.7     \\
\end{scotch}
    \label{tab:systematicLQ650}
\end{table*}

\section{Results of the leptoquark search}
\label{sec:results}

\begin{figure*}[hbtp!]
  \centering
    {\includegraphics[width=.49\textwidth]{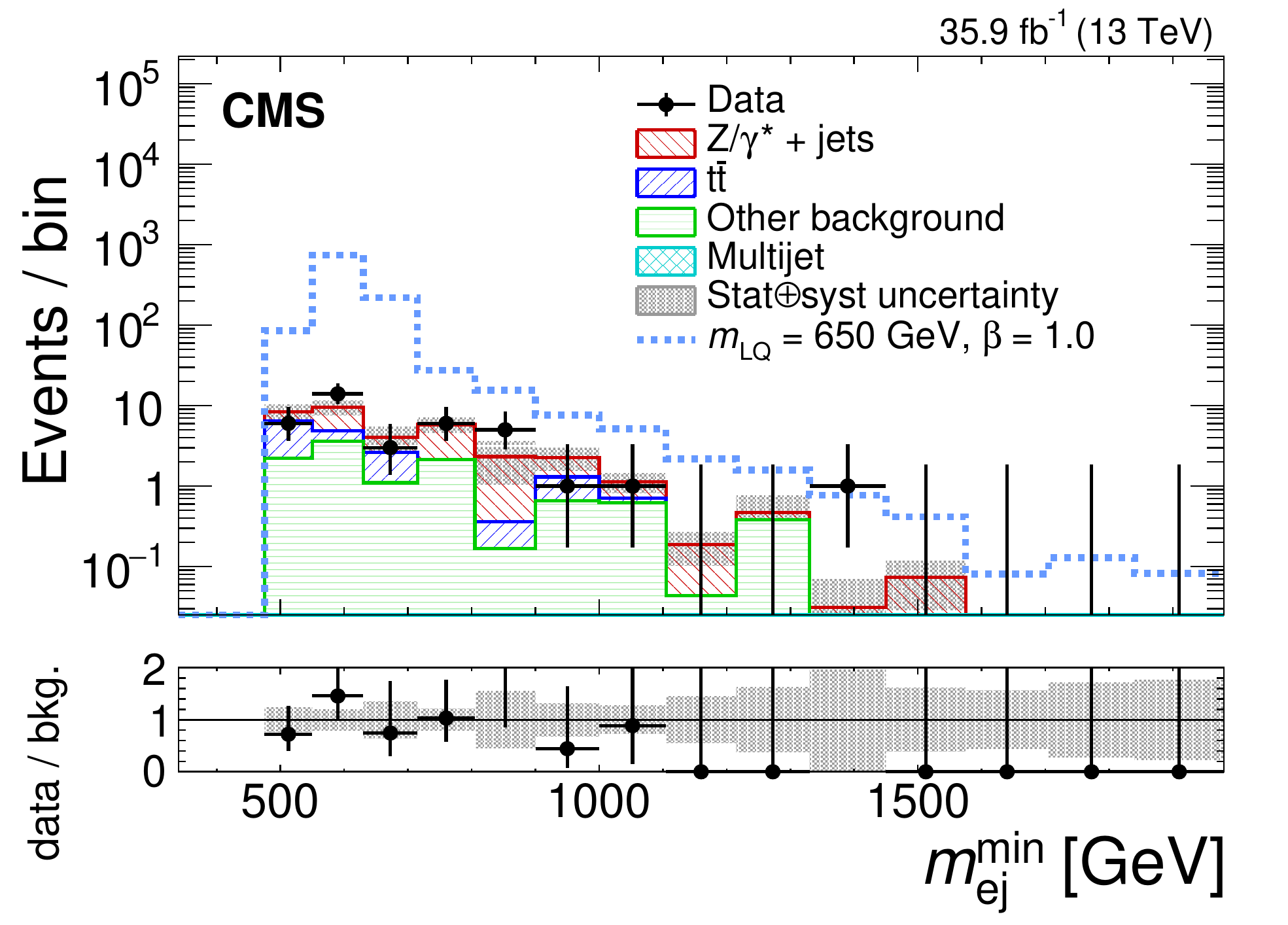}}
    {\includegraphics[width=.49\textwidth]{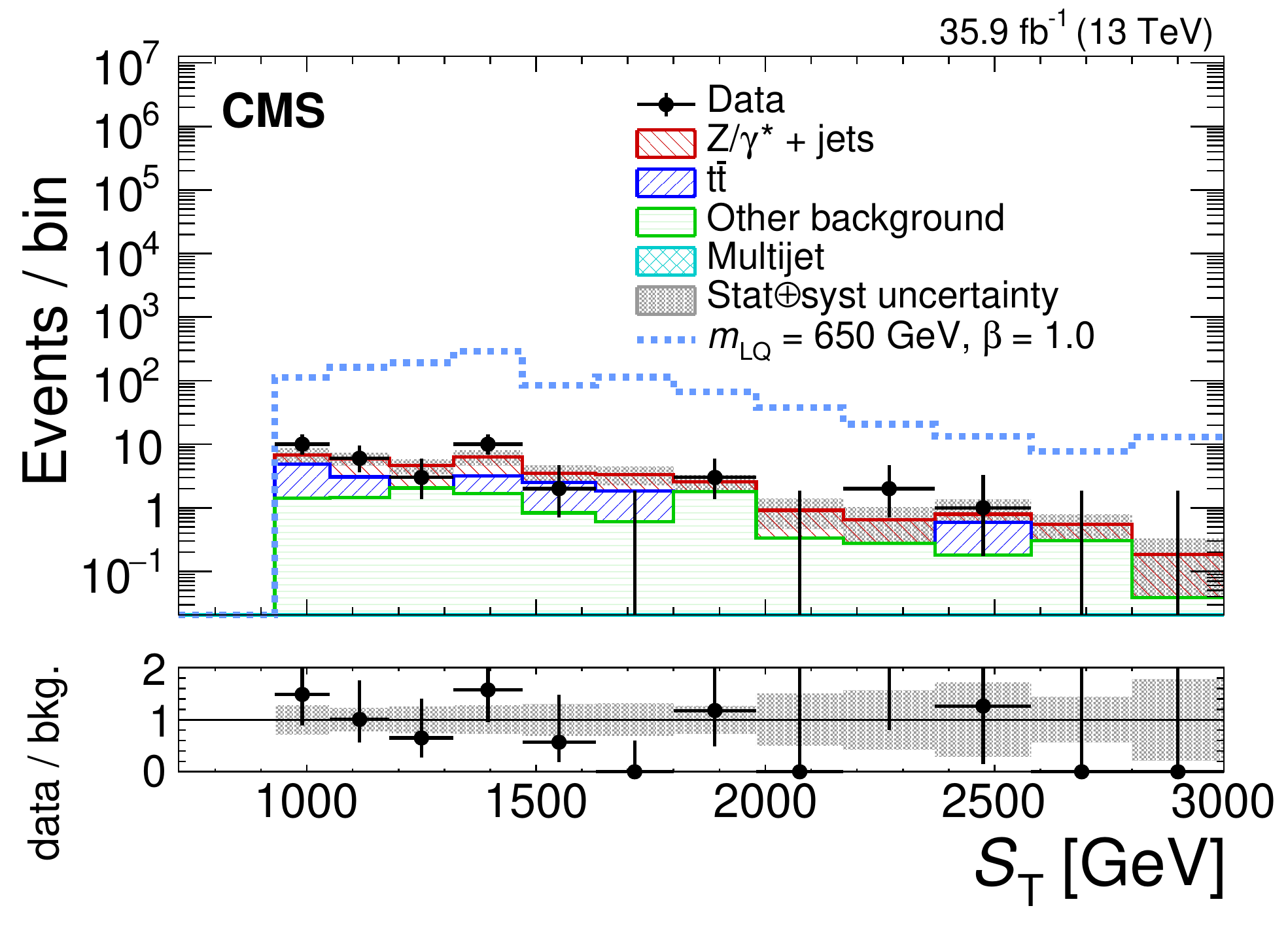}}
    {\includegraphics[width=.49\textwidth]{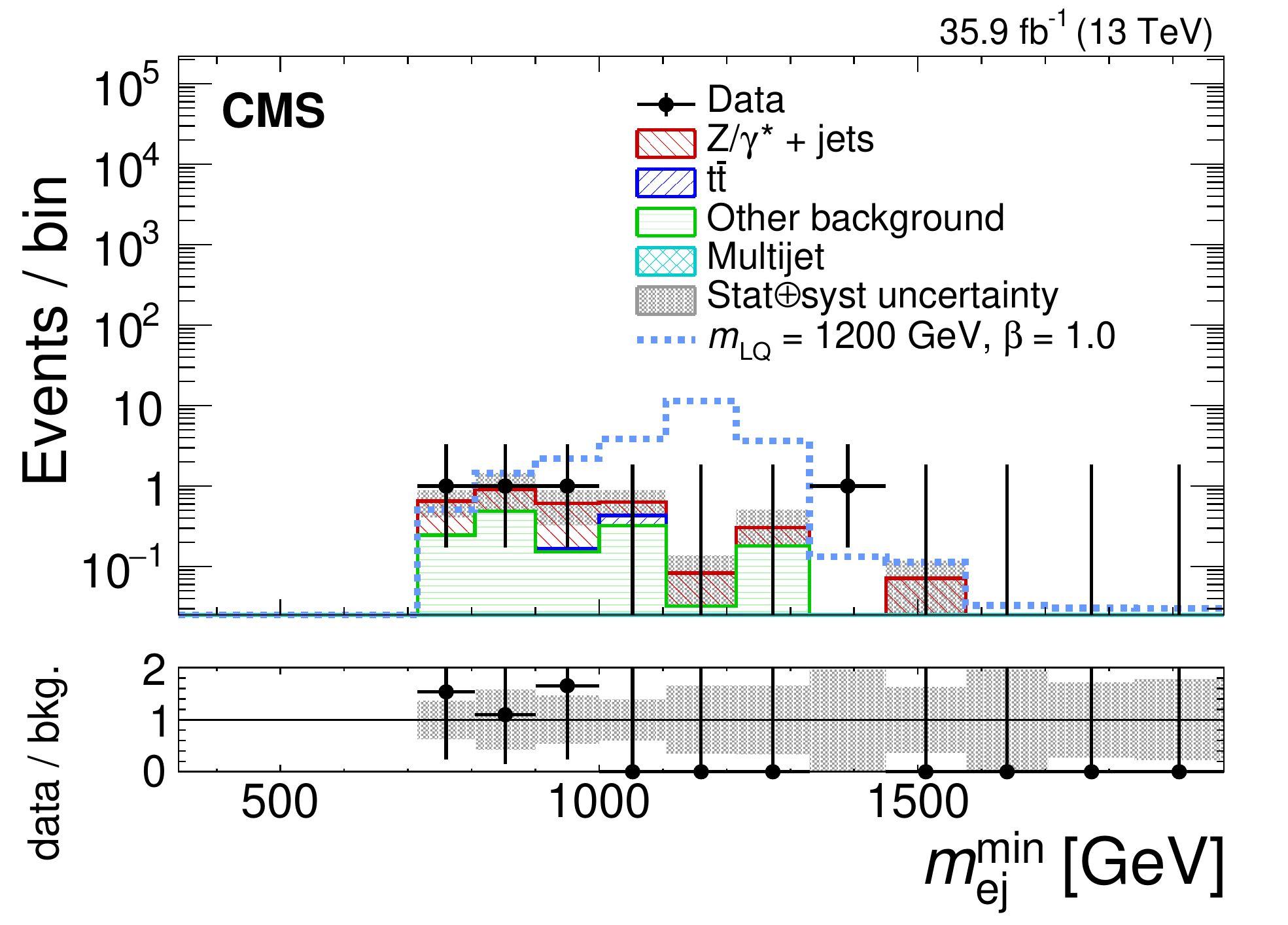}}
    {\includegraphics[width=.49\textwidth]{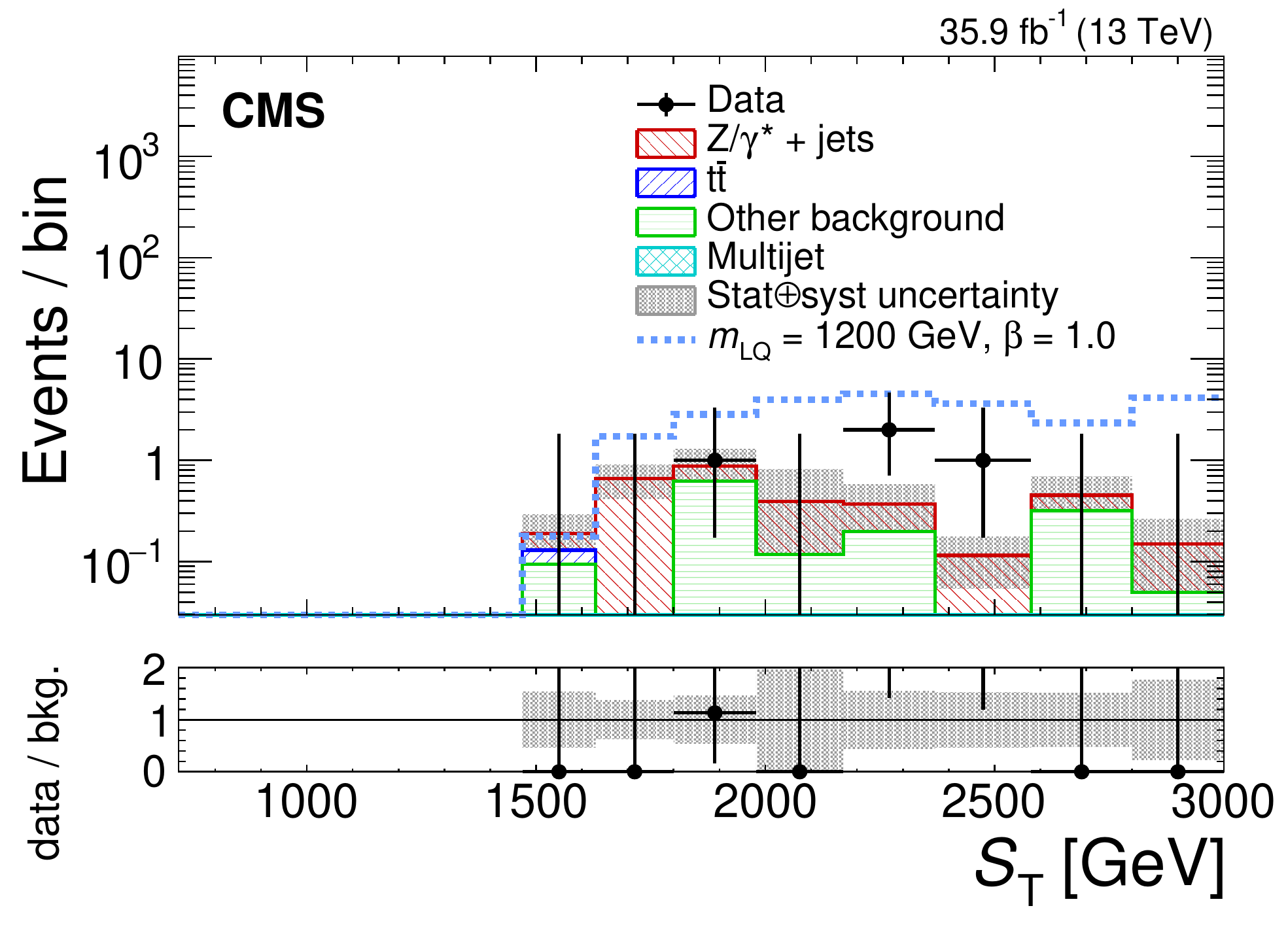}}
    \caption{\mejmin\,(left) and \st\,(right) distributions for events passing
    the \eejj\ final selection for LQs of mass 650\,(upper) and 1200\,(lower)\GeV.
    The predicted signal model distributions are shown, along with major backgrounds
    and ``other background'' which consists of the sum of the \wjets, diboson, single
    top quark, and \gammajets\ contributions.
    The background contributions are stacked, while the signal distributions are plotted unstacked.
    The dark shaded region indicates the statistical and systematic uncertainty in the total background.
    The last bin includes all events beyond the upper $x$-axis boundary.
    }
    \label{fig:eejjFinalSels}
\end{figure*}

\begin{figure*}[hbt!]
  \centering
    {\includegraphics[width=.49\textwidth]{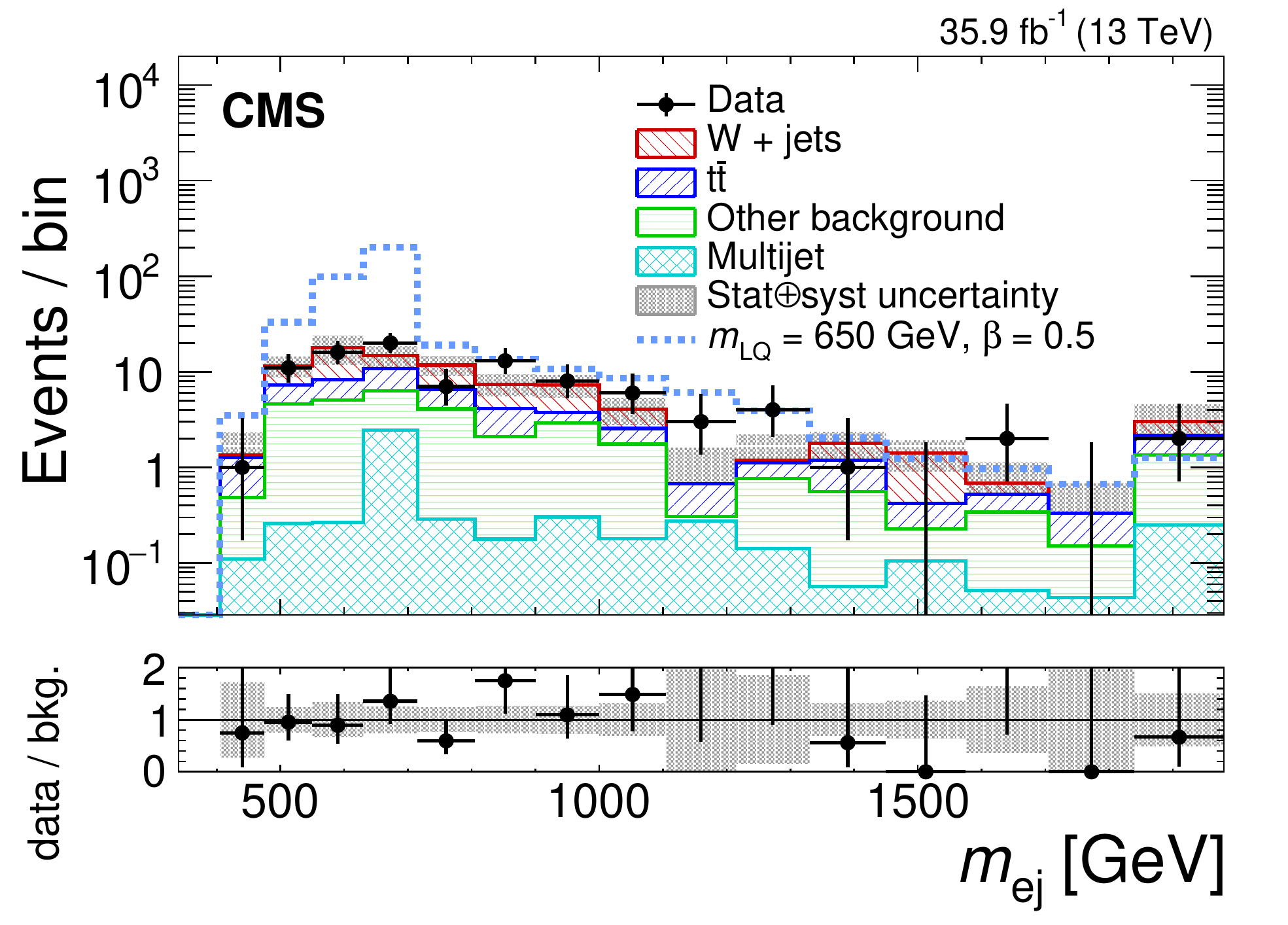}}
    {\includegraphics[width=.49\textwidth]{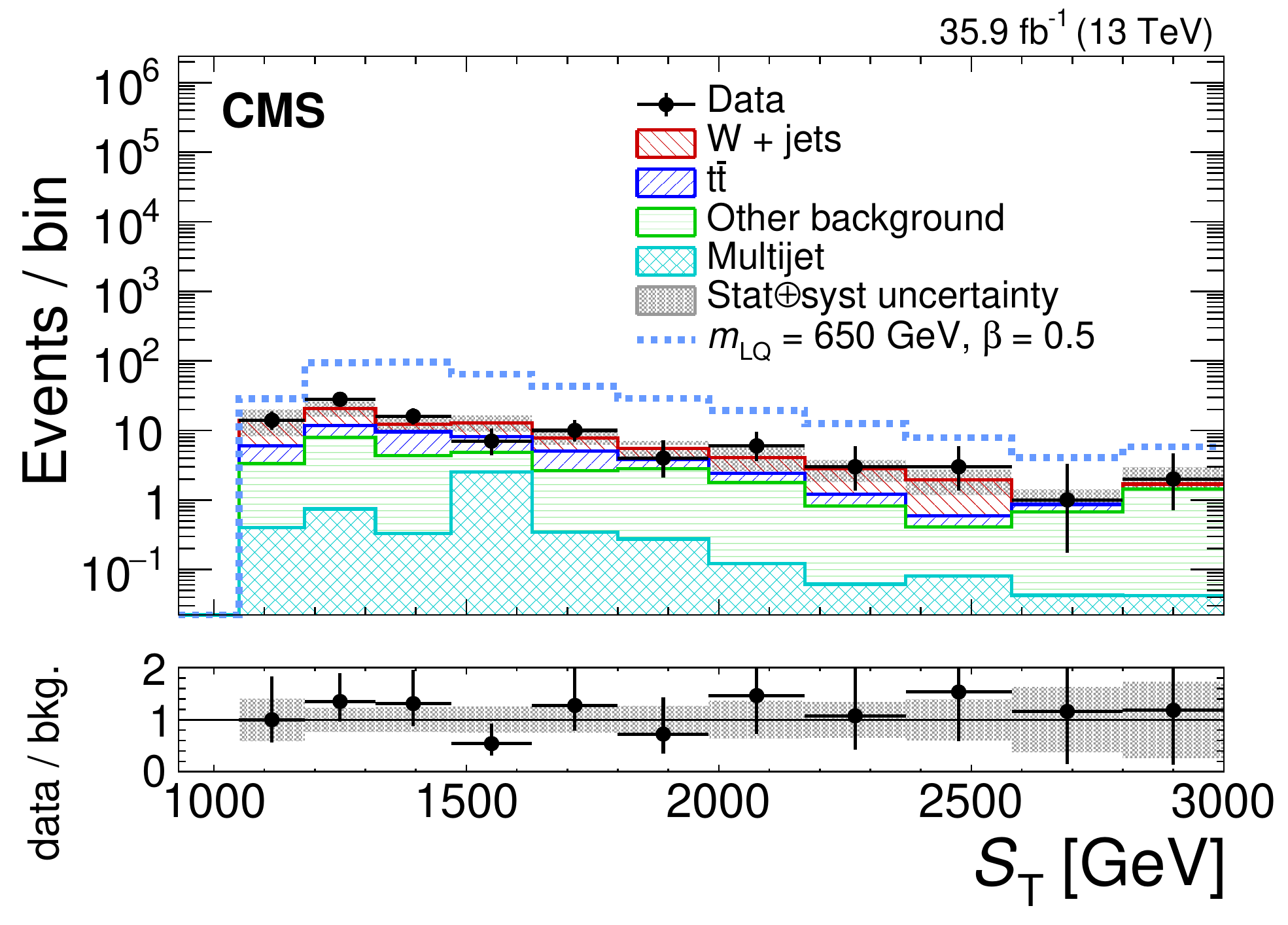}}
    {\includegraphics[width=.49\textwidth]{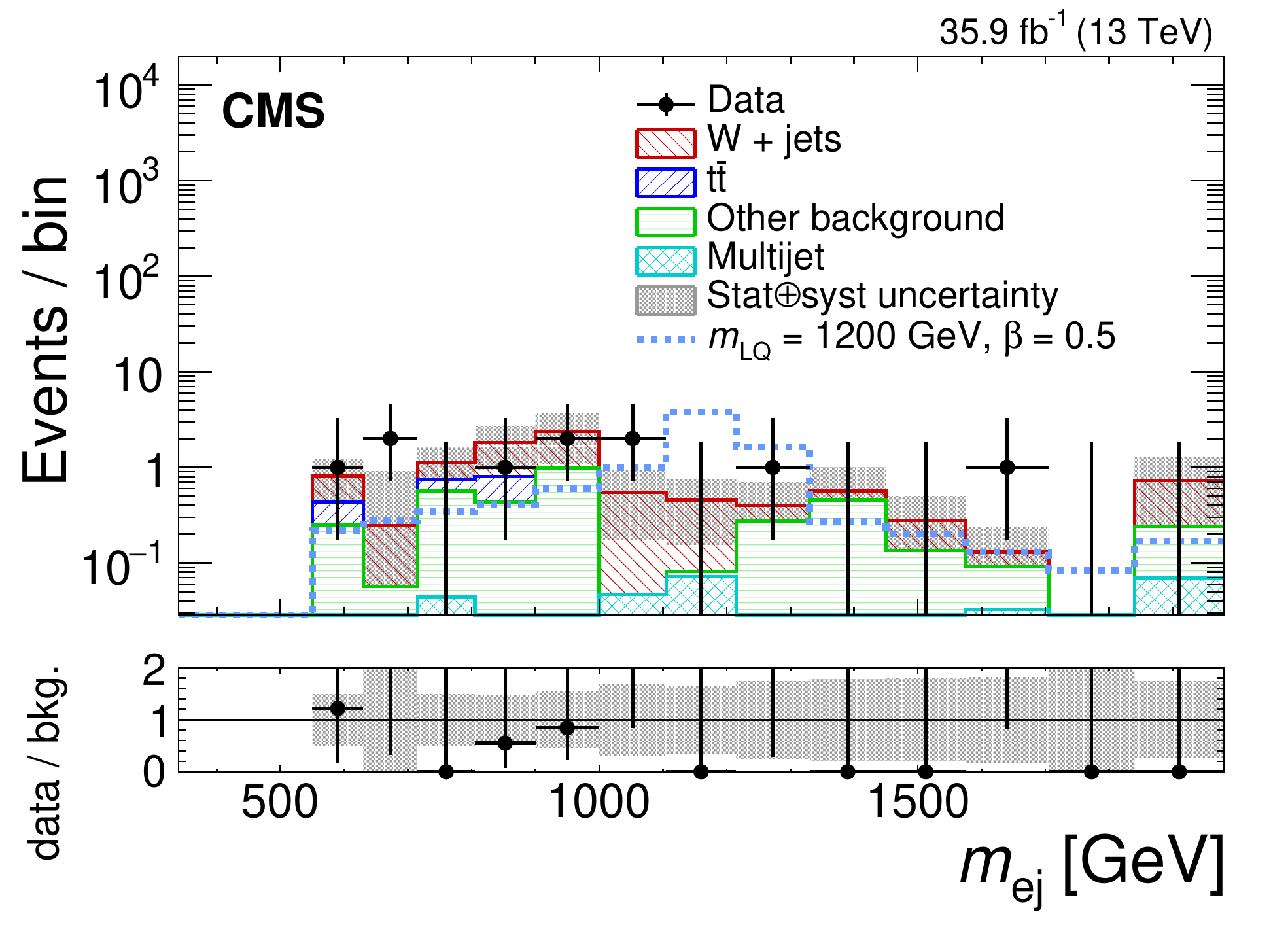}}
    {\includegraphics[width=.49\textwidth]{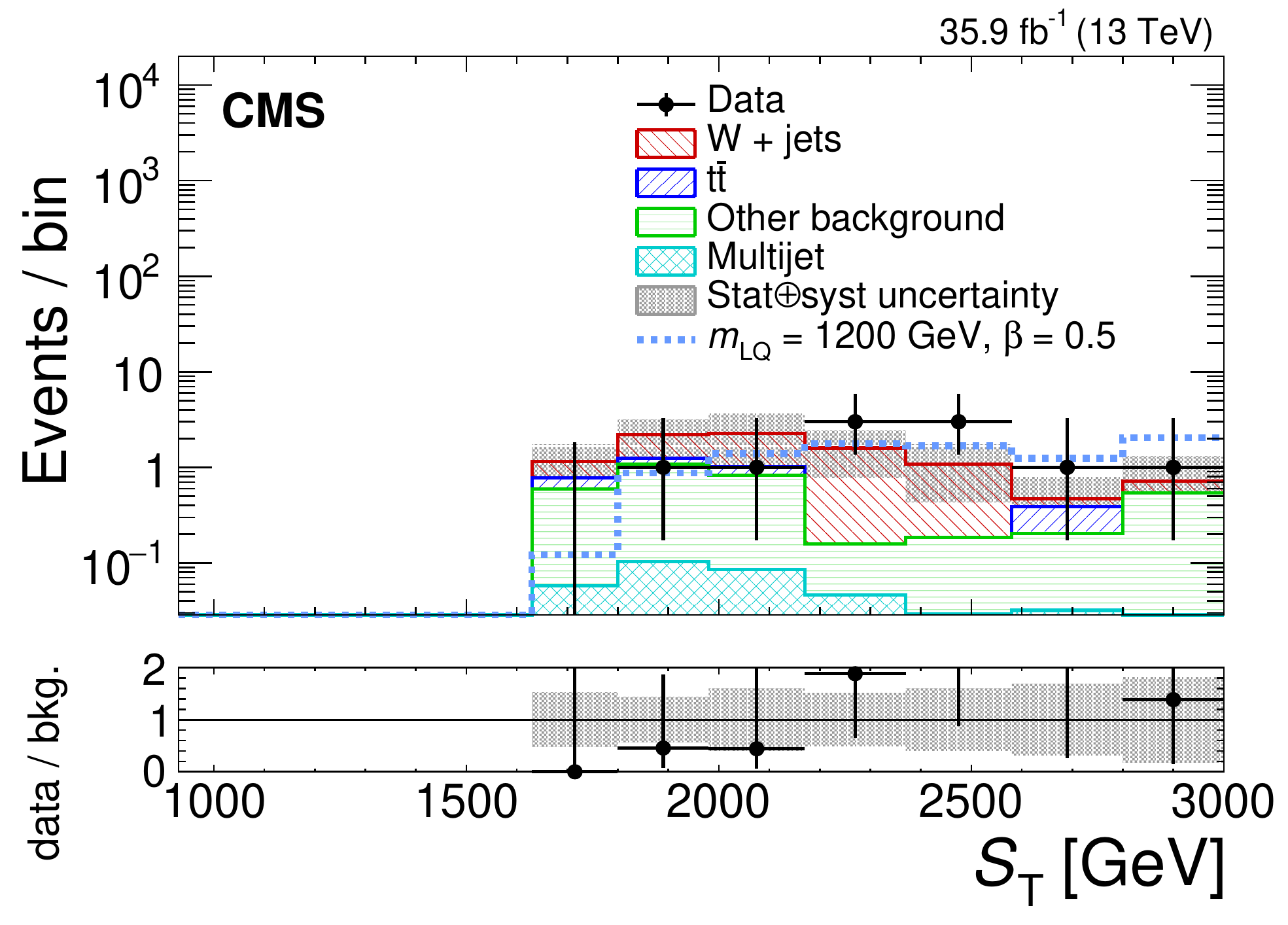}}
    \caption{\mej\,(left) and \st\,(right) distributions for events
      passing the \enujj\ final selection for LQs of mass
      650\,(upper) and 1200\,(lower)\GeV. The predicted signal model
      distributions are shown, along with major backgrounds and
      ``other background'' which consists of the sum of \zjets, diboson,
      single top quark, and \gammajets\ contributions. The background
      contributions are stacked, while the signal distributions are plotted
      unstacked. The dark shaded region indicates the statistical
      and systematic uncertainty in the total background. The last bin
      includes all events beyond the upper $x$-axis boundary.
    }
    \label{fig:enujjFinalSels}
\end{figure*}

After applying the final selection criteria shown in Fig.~\ref{fig:optimizationThresholds},
the data are compared to SM background expectations for both channels and
each \mlq hypothesis.
Distributions of \mejmin and \st are shown in
Fig.~\ref{fig:eejjFinalSels} for the \eejj channel with the selections
applied for the 650 and 1200\GeV \mlq hypotheses.
Figure~\ref{fig:enujjFinalSels} shows the corresponding distributions
of \mej\ and \st\ for the \enujj channel for the same mass hypotheses.

\begin{figure}[hbt!]
  \centering
    {\includegraphics[width=.49\textwidth]{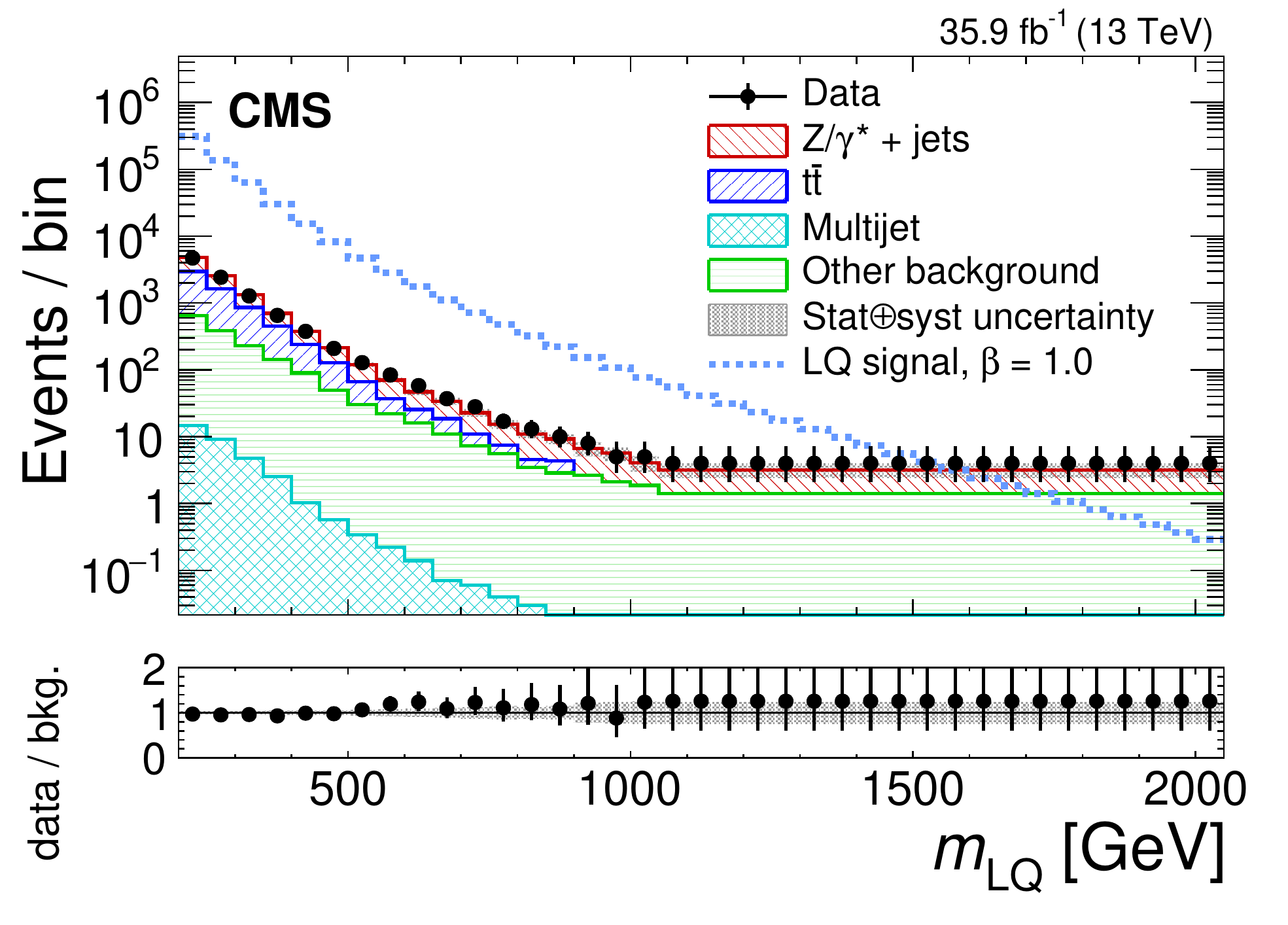}}
    {\includegraphics[width=.49\textwidth]{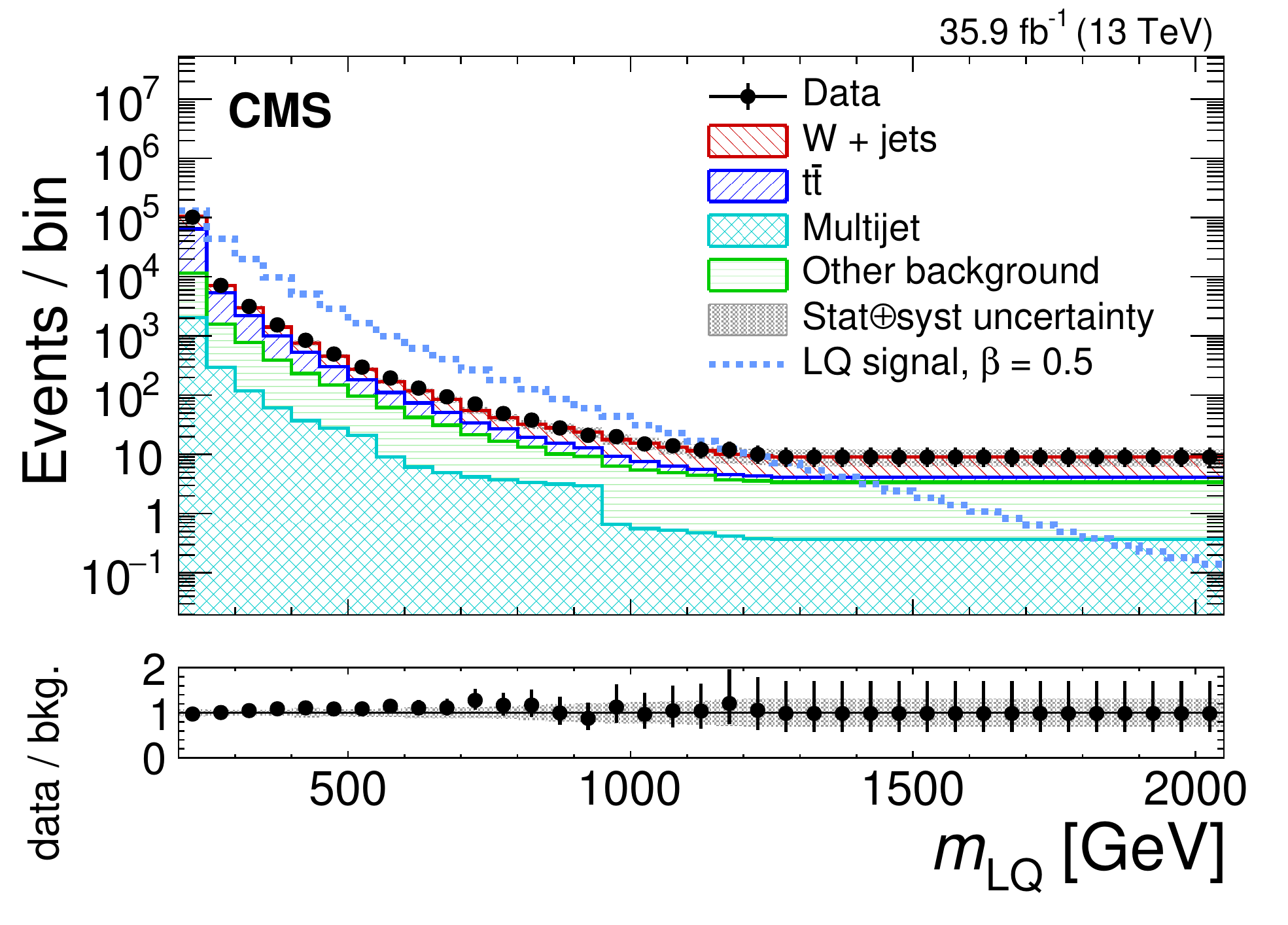}}
    \caption{
      Data, background, and expected signal yields after applying the final selection criteria for the \eejj (\cmsLeft) and \enujj (\cmsRight) channels.
      ``Other background" includes diboson, single top quark,
      and \wjets (for the \eejj channel) or \zjets (for the \enujj channel). The bin contents are correlated, because events selected for higher-mass
      LQ searches are a subset of those selected for lower mass searches.
    }
    \label{fig:finalEventYields}
\end{figure}

Figure~\ref{fig:finalEventYields} shows background, data,
and expected signal for each LQ mass point after applying the final selection criteria.
Signal efficiency times acceptance, along with tables listing event yields for
signal, background, and data are provided in Appendix~\ref{app:suppMat}.
The data are found to be in agreement with SM background expectations
in both channels.  We set upper limits on the product of the cross section
and branching fraction for scalar LQs as a function of \mlq and $\beta$.
The limits are calculated using the asymptotic approximation~\cite{Cowan:2010js}
of the \CLs  modified frequentist
approach~\cite{Junk:1999kv,Read:2002hq,CMS-NOTE-2011-005}.
Systematic uncertainties described in
Sec.~\ref{sec:systematics} are modeled with log-normal probability
density functions, while statistical uncertainties are modeled with
gamma functions whose widths are calculated from the number of events
in the control regions or simulated samples.

\begin{figure}[hbt]
  \includegraphics[width=0.49\textwidth]{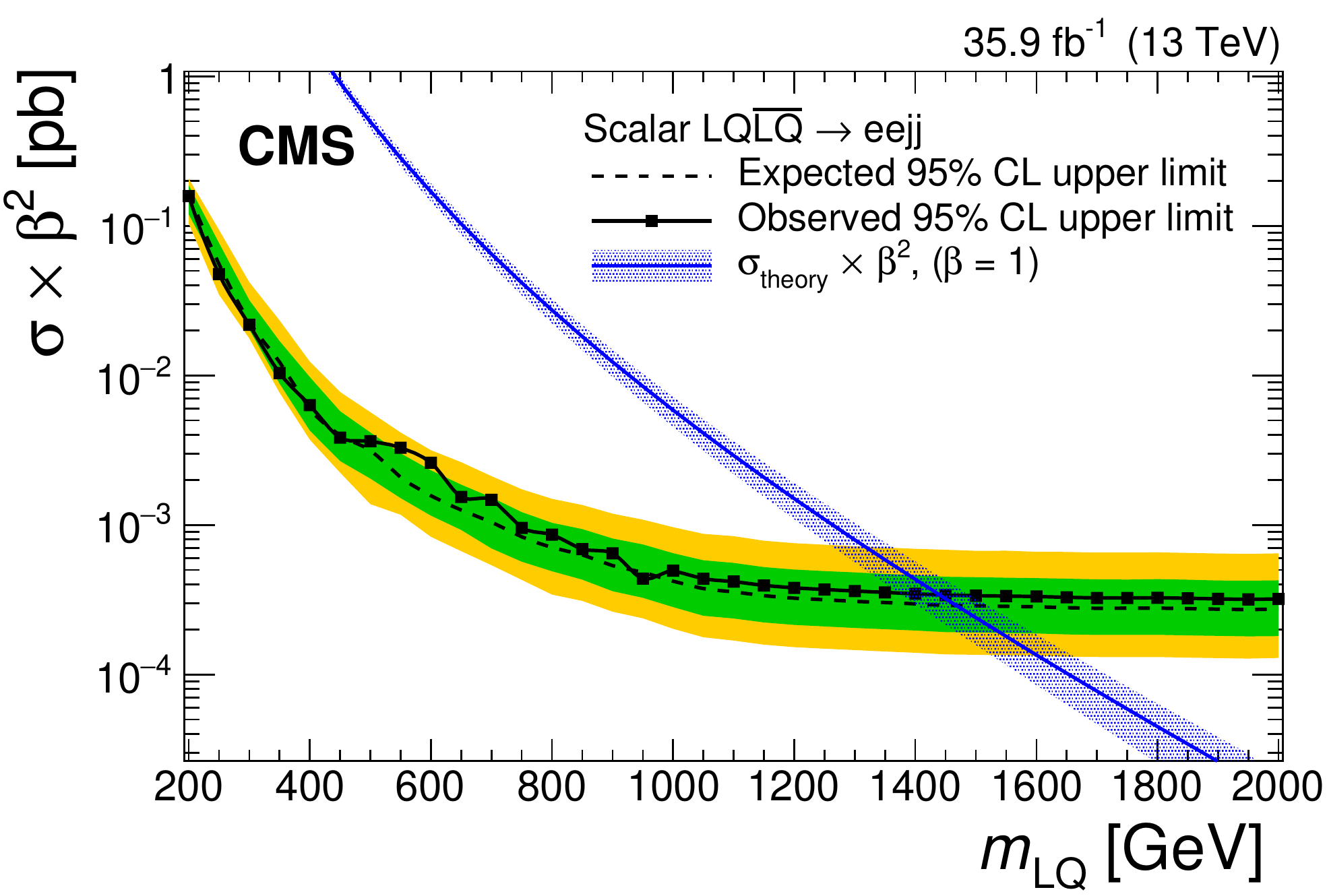}
  \includegraphics[width=0.49\textwidth]{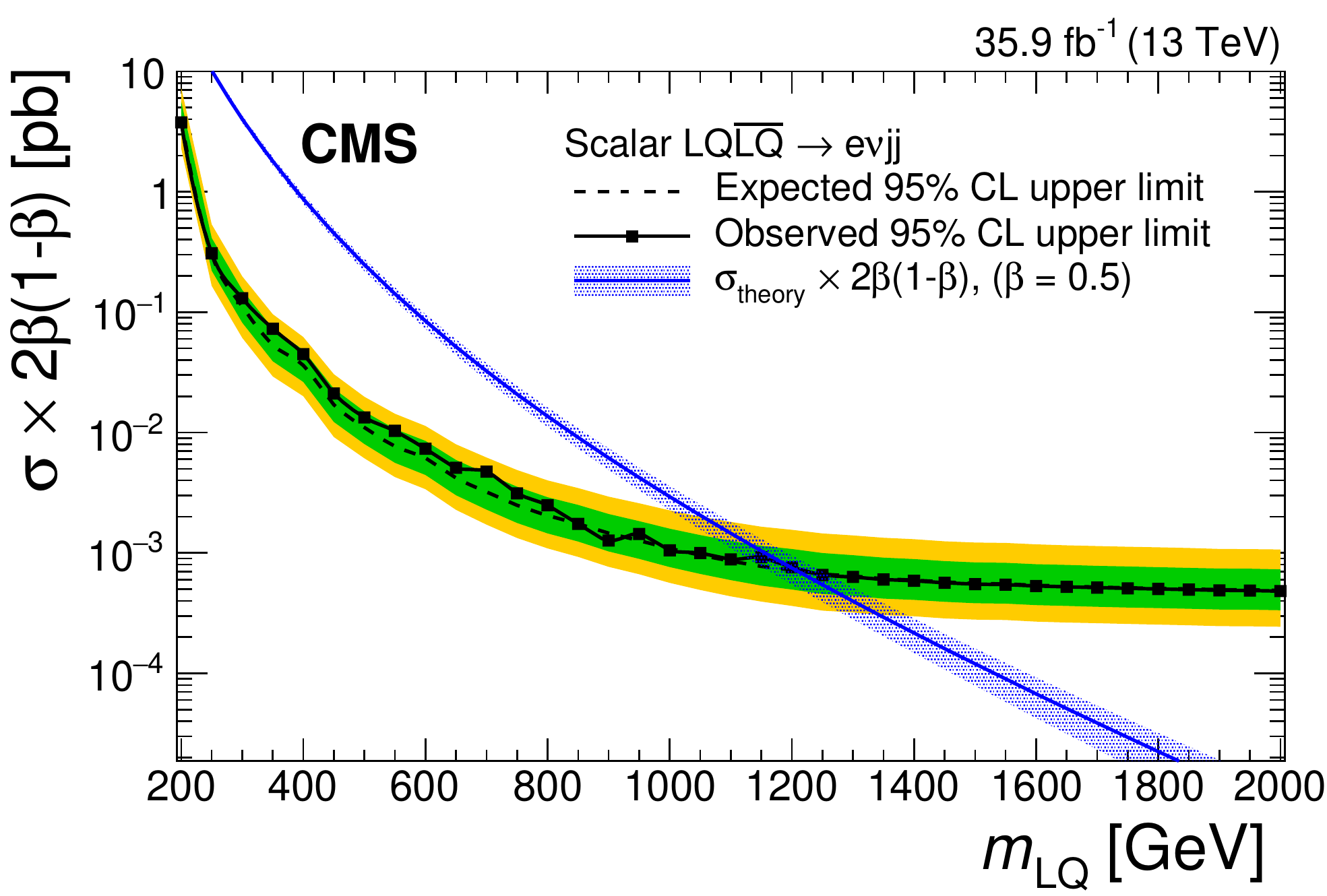}
  \caption{
    Observed upper limits for scalar LQ pair-production cross
    section times $\beta^2$ (\cmsLeft) and $\beta(1-\beta)$
    (\cmsRight) at 95\% \CL obtained with the \eejj (\cmsLeft)
    and \enujj (\cmsRight) analysis.  The median (dashed line),
    68\% (inner green band) and 95\% (outer yellow band)
    confidence-interval expected limits are also shown.
  }
  \label{fig:scalarLQLimits}
\end{figure}

We set upper limits on the production cross section multiplied by the
branching fraction $\beta^2$ or $2\beta(1-\beta)$ at 95\% \CL as a
function of \mlq. The expected and observed limits are shown with
NLO predictions for the scalar LQ pair production cross section
in Fig.~\ref{fig:scalarLQLimits} for both \eejj\ and \enujj\ channels.
The observed limits are within two standard deviations of expectations
from the background-only hypothesis. The uncertainty in the theoretical
prediction for the LQ pair production cross section is calculated as the
quadrature sum of the PDF uncertainty in the signal cross section and
the uncertainty due to the choice of renormalization and factorization
scales. The latter is estimated by independently varying the scales up
and down by a factor of two.

\begin{figure}[hbt]
  \centering
  \includegraphics[width=0.49\textwidth]{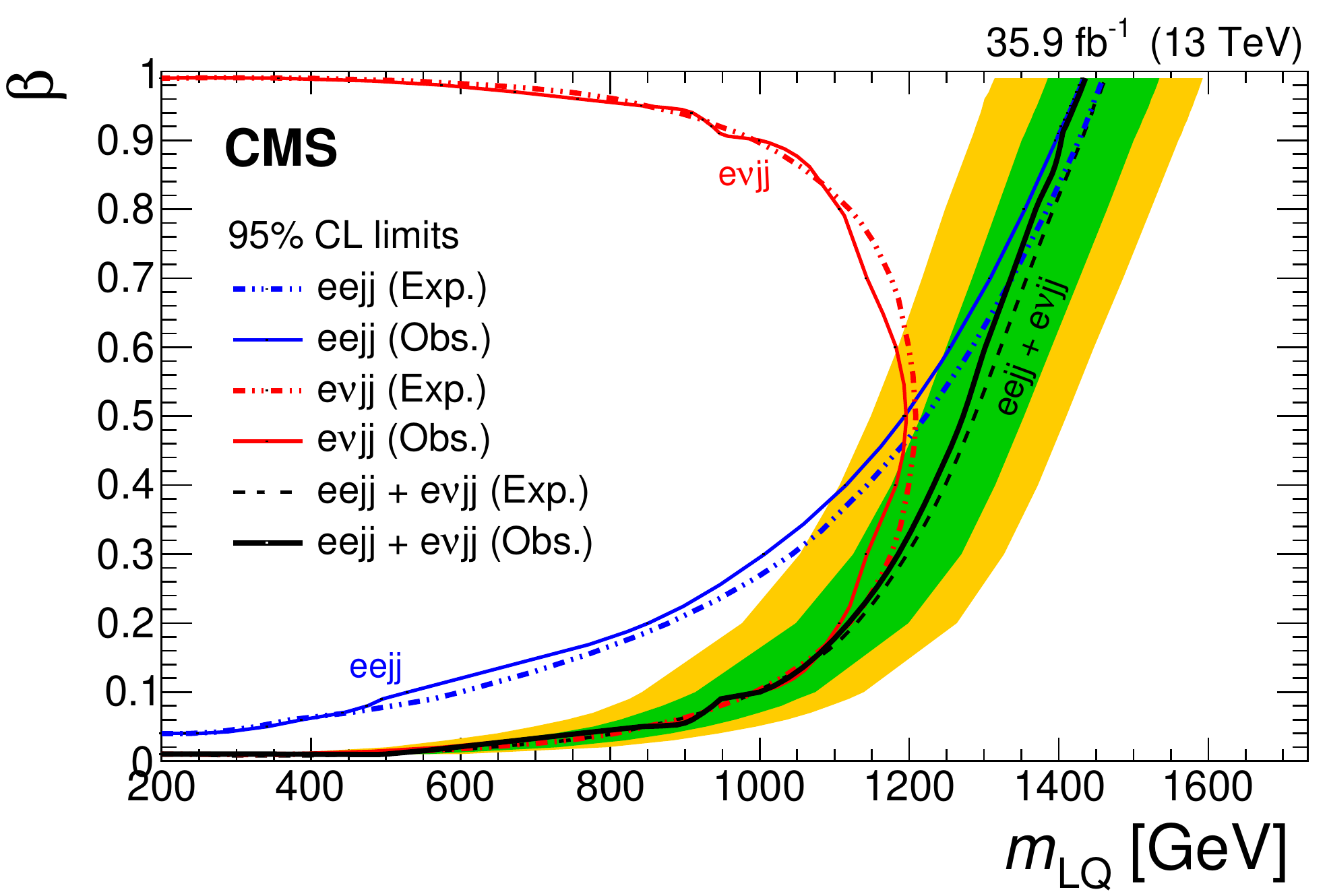}
  \caption{
    Expected and observed exclusion limits at 95\% \CL for pair production
    of first-generation scalar LQ shown in the $\beta$ versus
    \mlq\ plane for the individual \eejj and \enujj channels and
    their combination. The inner green and outer yellow bands represent
    the 68\% and 95\% confidence intervals on the expected limits.
  }
  \label{fig:Scan_limit}
\end{figure}

Under the assumption $\beta=1.0$, where only the \eejj\ channel
contributes, first-generation scalar LQs with masses less than
1435\GeV are excluded at 95\% \CL compared to a median expected
limit of 1465\GeV.
For $\beta=0.5$, using the \enujj channel alone, LQ masses are
excluded below 1195\GeV with the corresponding expected limit
being 1210\GeV. As both \eejj and \enujj decays contribute at
$\beta$ values smaller than 1, the LQ mass limit is improved
using the combination of the two channels. In this combination,
systematic uncertainties are considered to be fully correlated
between the channels, while statistical uncertainties are treated
as fully uncorrelated.
Limits for a range of $\beta$ values from 0 to 1 are set at 95\%
\CL for both \eejj and \enujj channels, as well as for their
combination, as shown in Fig.~\ref{fig:Scan_limit}.
In the $\beta=0.5$ case, the combination excludes first-generation
scalar LQs with masses less than 1270\GeV, compared to a median
expected value of 1285\GeV.

\section{\texorpdfstring{$R$}{R}-parity violating supersymmetry interpretation}
\label{sec:RPVSUSY}

Many new physics models predict the existence of particles with
couplings of the type expected for LQs. One such model is $R$-parity
violating supersymmetry (RPV SUSY)~\cite{rpv,Barbier:2004ez}, where
the superpartners of quarks or `squarks' can decay into LQ-like
final states. For example, the top squark (\PSQt) can decay to a
bottom quark and an electron. The topology of the resulting events
is similar to an LQ decay and hence these events will pass our
nominal selection for the \eejj channel.

The analysis is recast in terms of the possible production of
prompt top-squark pairs ($c\tau=0\unit{cm}$), with each \PSQt
subsequently decaying to a bottom quark and an electron.
Limits on the production cross section for
\PSQt\ pairs are calculated from the \eejj data, accounting for
the difference in branching fractions of LQ and \PSQt
decays to electrons.

\begin{figure}[hbt]
  \centering
  \includegraphics[width=0.49\textwidth]{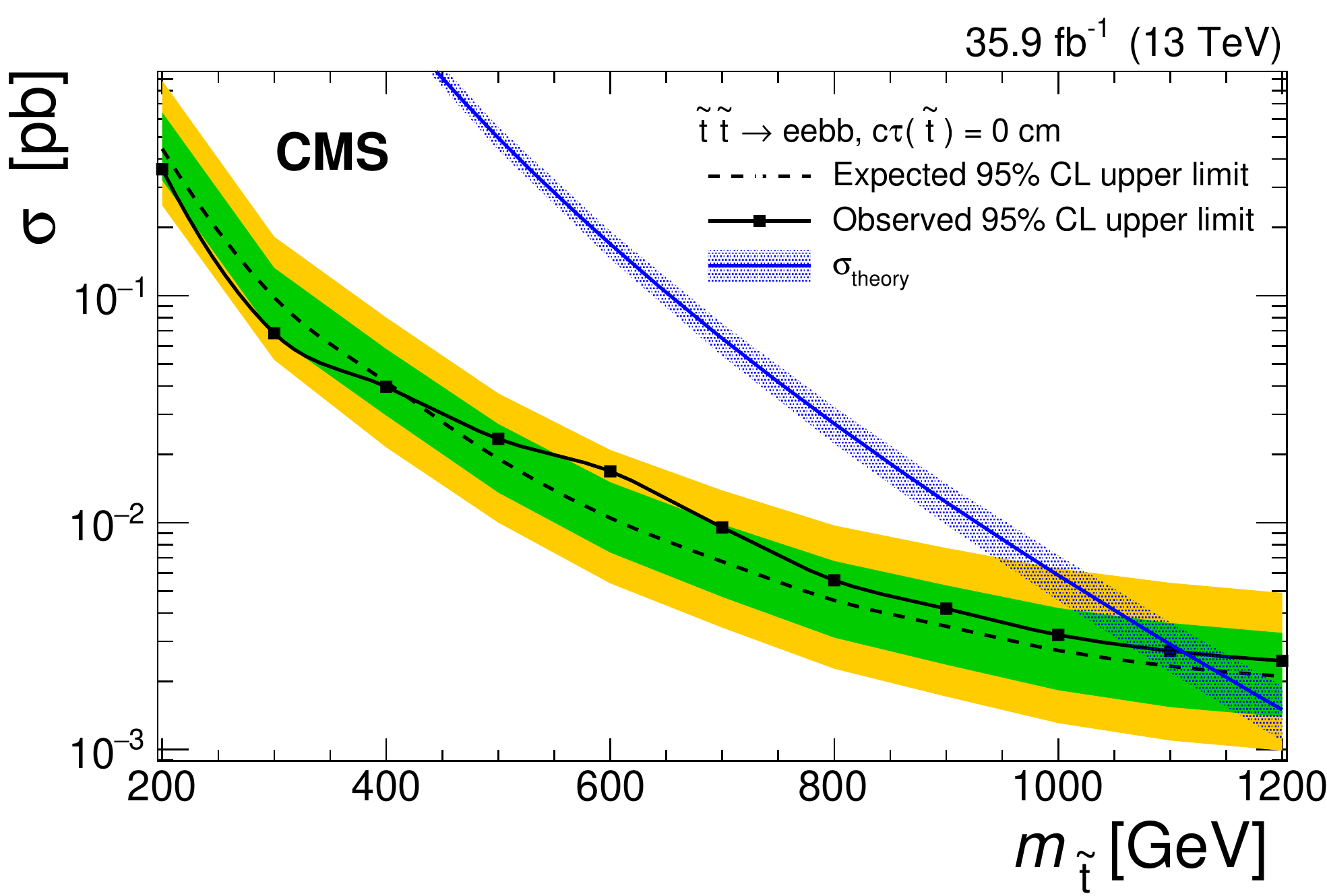}
  \caption{Expected and observed upper limits at 95\% \CL on the RPV
    SUSY \PSQt squark pair production cross section as a function of
    $M_{\PSQt}$ for $c\tau=0$\unit{cm}. The expected limits represent
    the median values, while the inner green and outer yellow bands
    are the 68\% and 95\% confidence intervals, respectively.}
  \label{fig:rpv-susy}
\end{figure}

Figure~\ref{fig:rpv-susy} shows the expected and observed 95\% \CL
upper limits on the RPV SUSY \PSQt pair production cross section as a
function of the \PSQt squark mass ($m_{\PSQt}$). The observed exclusion
limit is 1100\GeV for $c\tau=0$\unit{cm}.

\section{Summary}
\label{sec:summary}

A search has been performed for the pair production of first-generation
scalar leptoquarks in final states consisting of two high-momentum
electrons and two jets, or one electron, large missing transverse
momentum and two jets.  The data sample used in the study corresponds to
an integrated luminosity of 35.9\fbinv recorded by the CMS experiment
at $\sqrt{s}=13\TeV$.  The data are found to be in agreement with standard
model background expectations and a lower limit at 95\% confidence level
is set on the scalar leptoquark mass at 1435\,(1270)\GeV for
$\beta=$ 1.0\,(0.5), where $\beta$ is the branching fraction of the
leptoquark decay to an electron and a quark.
These results constitute the most stringent limits on the mass of first-generation
scalar leptoquarks to date.  The data are also interpreted in the context of
an $R$-parity violating supersymmetric model with promptly decaying top squarks,
which can decay into leptoquark-like final states.  Top squarks
are excluded for masses below 1100\GeV.

\begin{acknowledgments}
We congratulate our colleagues in the CERN accelerator departments for the excellent performance of the LHC and thank the technical and administrative staffs at CERN and at other CMS institutes for their contributions to the success of the CMS effort. In addition, we gratefully acknowledge the computing centers and personnel of the Worldwide LHC Computing Grid for delivering so effectively the computing infrastructure essential to our analyses. Finally, we acknowledge the enduring support for the construction and operation of the LHC and the CMS detector provided by the following funding agencies: BMBWF and FWF (Austria); FNRS and FWO (Belgium); CNPq, CAPES, FAPERJ, FAPERGS, and FAPESP (Brazil); MES (Bulgaria); CERN; CAS, MoST, and NSFC (China); COLCIENCIAS (Colombia); MSES and CSF (Croatia); RPF (Cyprus); SENESCYT (Ecuador); MoER, ERC IUT, and ERDF (Estonia); Academy of Finland, MEC, and HIP (Finland); CEA and CNRS/IN2P3 (France); BMBF, DFG, and HGF (Germany); GSRT (Greece); NKFIA (Hungary); DAE and DST (India); IPM (Iran); SFI (Ireland); INFN (Italy); MSIP and NRF (Republic of Korea); MES (Latvia); LAS (Lithuania); MOE and UM (Malaysia); BUAP, CINVESTAV, CONACYT, LNS, SEP, and UASLP-FAI (Mexico); MOS (Montenegro); MBIE (New Zealand); PAEC (Pakistan); MSHE and NSC (Poland); FCT (Portugal); JINR (Dubna); MON, RosAtom, RAS, RFBR, and NRC KI (Russia); MESTD (Serbia); SEIDI, CPAN, PCTI, and FEDER (Spain); MOSTR (Sri Lanka); Swiss Funding Agencies (Switzerland); MST (Taipei); ThEPCenter, IPST, STAR, and NSTDA (Thailand); TUBITAK and TAEK (Turkey); NASU and SFFR (Ukraine); STFC (United Kingdom); DOE and NSF (USA).

\hyphenation{Rachada-pisek} Individuals have received support from the Marie-Curie program and the European Research Council and Horizon 2020 Grant, contract No. 675440 (European Union); the Leventis Foundation; the A. P. Sloan Foundation; the Alexander von Humboldt Foundation; the Belgian Federal Science Policy Office; the Fonds pour la Formation \`a la Recherche dans l'Industrie et dans l'Agriculture (FRIA-Belgium); the Agentschap voor Innovatie door Wetenschap en Technologie (IWT-Belgium); the F.R.S.-FNRS and FWO (Belgium) under the ``Excellence of Science - EOS" - be.h project n. 30820817; the Ministry of Education, Youth and Sports (MEYS) of the Czech Republic; the Lend\"ulet (``Momentum") Program and the J\'anos Bolyai Research Scholarship of the Hungarian Academy of Sciences, the New National Excellence Program \'UNKP, the NKFIA research grants 123842, 123959, 124845, 124850 and 125105 (Hungary); the Council of Science and Industrial Research, India; the HOMING PLUS program of the Foundation for Polish Science, cofinanced from European Union, Regional Development Fund, the Mobility Plus program of the Ministry of Science and Higher Education, the National Science Center (Poland), contracts Harmonia 2014/14/M/ST2/00428, Opus 2014/13/B/ST2/02543, 2014/15/B/ST2/03998, and 2015/19/B/ST2/02861, Sonata-bis 2012/07/E/ST2/01406; the National Priorities Research Program by Qatar National Research Fund; the Programa Estatal de Fomento de la Investigaci{\'o}n Cient{\'i}fica y T{\'e}cnica de Excelencia Mar\'{\i}a de Maeztu, grant MDM-2015-0509 and the Programa Severo Ochoa del Principado de Asturias; the Thalis and Aristeia programs cofinanced by EU-ESF and the Greek NSRF; the Rachadapisek Sompot Fund for Postdoctoral Fellowship, Chulalongkorn University and the Chulalongkorn Academic into Its 2nd Century Project Advancement Project (Thailand); the Welch Foundation, contract C-1845; and the Weston Havens Foundation (USA).

\end{acknowledgments}

\clearpage
\bibliography{auto_generated}

\appendix
\section{Efficiencies and event yields\label{app:suppMat}}
\begin{figure*}[htb]
  \includegraphics[width=.49\textwidth]{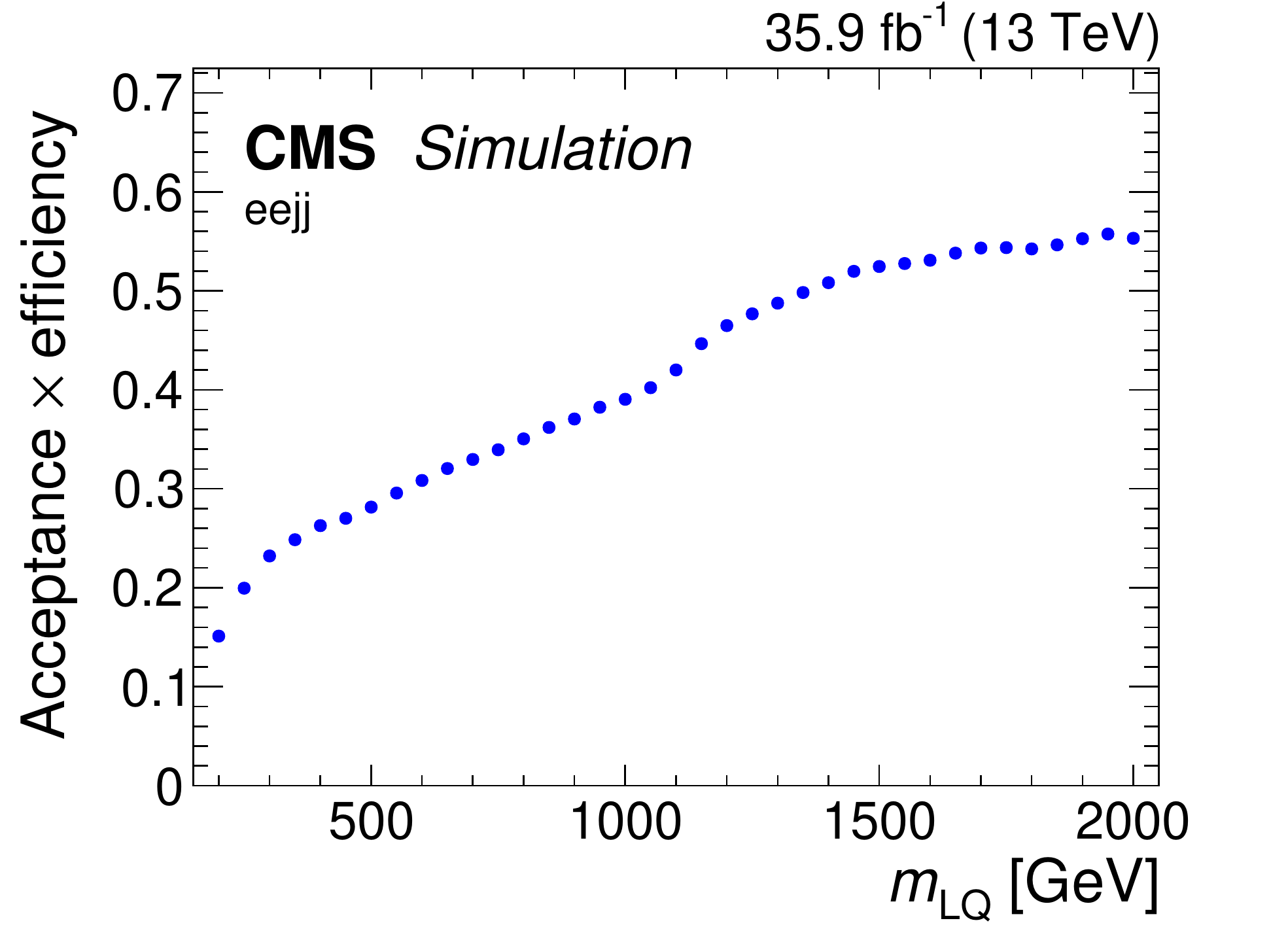}
  \includegraphics[width=.49\textwidth]{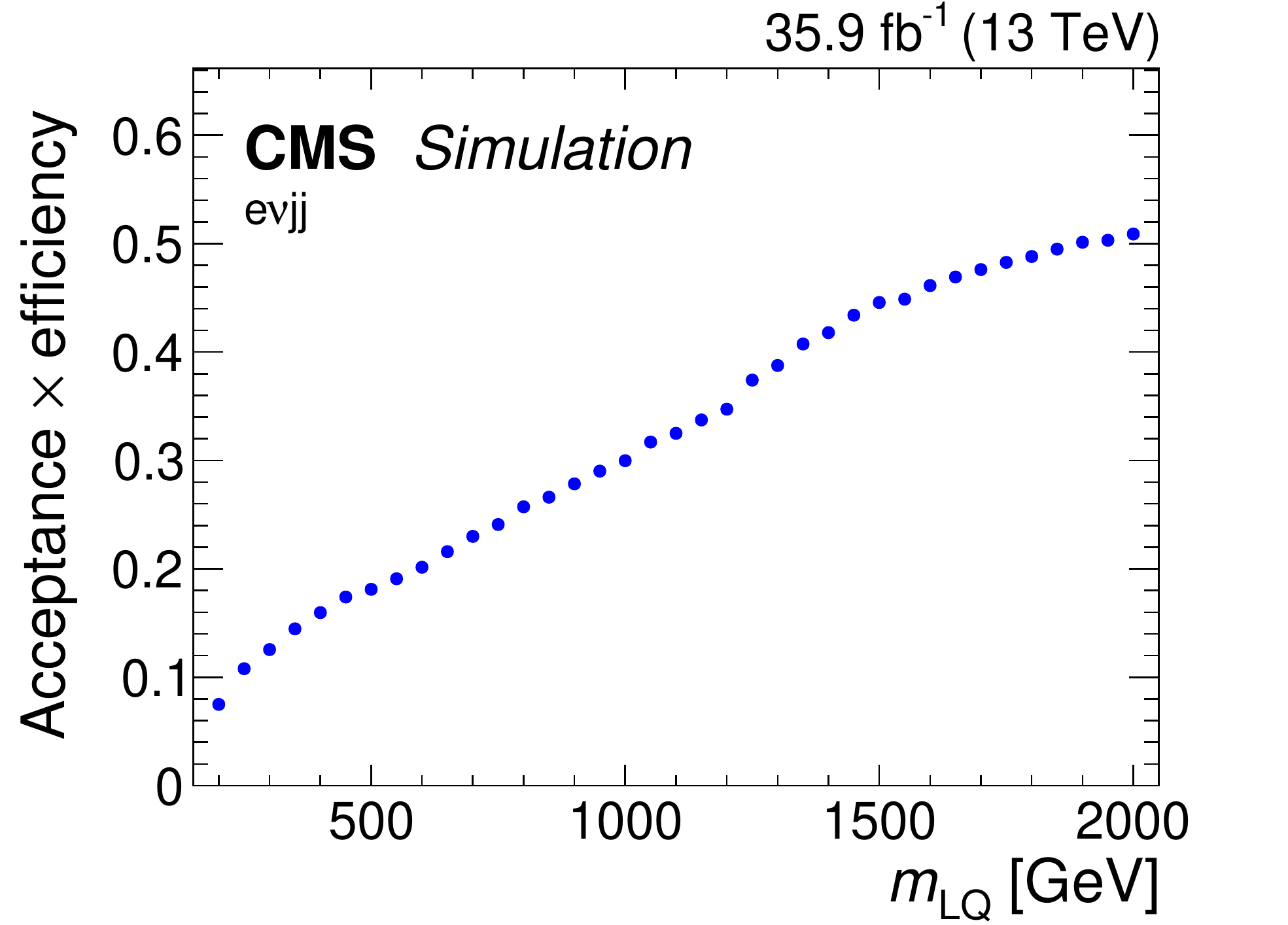}
  \caption{
    The product of signal acceptance and efficiency after final optimized selections,
    as a function of \mlq for the \eejj (left) and \enujj\ (right) channels.
  }
  \label{fig:sigAccEff}
\end{figure*}

In Fig.~\ref{fig:sigAccEff} the product of signal acceptance and efficiency is
shown after final optimized selections as a function of \mlq for the \eejj\,(left)
and \enujj\,(right) channels.
Tables~\ref{tab:eejjFinalSels} and \ref{tab:enujjFinalSels} list the
number of events passing the final selection criteria in data and the various
background components as a function of \mlq\
for the \eejj and \enujj\ channels, respectively.

\begin{table*}[htb]
  \centering
    \topcaption{Event yields after the optimized \eejj\ selections.
    Uncertainties are statistical except for the total background, where both
    statistical and systematic uncertainties are shown. An entry of $0.0$ quoted
    for the uncertainty indicates that its value is negligibly small. LQ masses
    are given in units of \GeV and init. sel. refers to initial selection.}
    \label{tab:eejjFinalSels}
    \def\arraystretch{1.5}
    \cmsTable{
    \begin{scotch}{cccccccc}

	LQ mass		&		Signal				&		\zjets				&		\ttbar							&		Multijet				&		\vv, \PW, single \cPqt, \gammajets							&		Total background									&		Data		\\ \hline
init. sel.		&	$	\NA			$	&	$	41600	\pm	49	$	&	$	7100	\pm	68				$	&	$	26	\pm	0.1	$	&	$	2400	\pm	36				$	&	$	51100	\pm	91				\pm	2700	$	&	$	50585	$	\\
\noalign{\vskip 2mm}
	200		&	$	311500	\pm	3300	$	&	$	1900	\pm	16	$	&	$	2300	\pm	39				$	&	$	15	\pm	0.1	$	&	$	630	\pm	18				$	&	$	4800	\pm	46				\pm	120	$	&	$	4709	$	\\
	250		&	$	137400	\pm	1200	$	&	$	910	\pm	11	$	&	$	1200	\pm	29				$	&	$	9.1	\pm	0.1	$	&	$	380	\pm	14				$	&	$	2500	\pm	34				\pm	69	$	&	$	2426	$	\\
	300		&	$	63160	\pm	510	$	&	$	470	\pm	4.2	$	&	$	630	\pm	22				$	&	$	4.8	\pm	0.0	$	&	$	220	^{	+10	}_{	-9.5	}	$	&	$	1300	^{	+24	}_{	-24	}	\pm	24	$	&	$	1278	$	\\
	350		&	$	30150	\pm	230	$	&	$	250	\pm	2.7	$	&	$	310	\pm	15				$	&	$	2.5	\pm	0.0	$	&	$	140	^{	+9.5	}_{	-8.6	}	$	&	$	700	^{	+18	}_{	-18	}	\pm	27	$	&	$	652	$	\\
	400		&	$	15440	\pm	110	$	&	$	140	\pm	1.8	$	&	$	150	\pm	11				$	&	$	1.0	\pm	0.0	$	&	$	89	^{	+7.2	}_{	-6.2	}	$	&	$	380	^{	+13	}_{	-13	}	\pm	11	$	&	$	376	$	\\
	450		&	$	8260	\pm	60	$	&	$	85	\pm	1.5	$	&	$	79	\pm	7.7				$	&	$	0.6	\pm	0.0	$	&	$	49	^{	+2.3	}_{	-2.3	}	$	&	$	210	^{	+8.2	}_{	-8.1	}	\pm	5.3	$	&	$	209	$	\\
	500		&	$	4700	\pm	33	$	&	$	54	\pm	1.1	$	&	$	36	\pm	5.5				$	&	$	0.3	\pm	0.0	$	&	$	30	^{	+2.0	}_{	-1.9	}	$	&	$	120	^{	+6.0	}_{	-5.9	}	\pm	4.4	$	&	$	128	$	\\
	550		&	$	2830	\pm	19	$	&	$	33	\pm	0.8	$	&	$	15	\pm	4.0				$	&	$	0.2	\pm	0.0	$	&	$	22	^{	+1.8	}_{	-1.8	}	$	&	$	70	^{	+4.5	}_{	-4.5	}	\pm	2.6	$	&	$	84	$	\\
	600		&	$	1750	\pm	12	$	&	$	21	\pm	0.6	$	&	$	9.6	\pm	3.3				$	&	$	0.1	\pm	0.0	$	&	$	16	^{	+1.6	}_{	-1.6	}	$	&	$	47	^{	+3.7	}_{	-3.7	}	\pm	1.9	$	&	$	58	$	\\
	650		&	$	1110	\pm	7.2	$	&	$	15	\pm	0.6	$	&	$	7.7	\pm	2.9				$	&	$	0.1	\pm	0.0	$	&	$	11	^{	+1.4	}_{	-1.3	}	$	&	$	34	^{	+3.2	}_{	-3.2	}	\pm	1.3	$	&	$	37	$	\\
	700		&	$	718	\pm	4.5	$	&	$	12	\pm	0.5	$	&	$	3.7	\pm	2.2				$	&	$	0.1	\pm	0.0	$	&	$	7.3	^{	+1.2	}_{	-1.2	}	$	&	$	23	^{	+2.6	}_{	-2.6	}	\pm	1.0	$	&	$	28	$	\\
	750		&	$	470	\pm	2.9	$	&	$	7.8	\pm	0.3	$	&	$	2.0	\pm	1.9				$	&	$	0.0	\pm	0.0	$	&	$	5.5	^{	+1.1	}_{	-1.1	}	$	&	$	15	^{	+2.2	}_{	-2.2	}	\pm	0.6	$	&	$	17	$	\\
	800		&	$	320	\pm	1.9	$	&	$	6.4	\pm	0.4	$	&	$	1.1	^{	+0.5	}_{	-0.4	}	$	&	$	0.0	\pm	0.0	$	&	$	3.5	^{	+1.1	}_{	-0.9	}	$	&	$	11	^{	+1.2	}_{	-1.1	}	\pm	0.6	$	&	$	13	$	\\
	850		&	$	220	\pm	1.3	$	&	$	4.9	\pm	0.3	$	&	$	1.5	^{	+0.7	}_{	-0.5	}	$	&	$	0.0	\pm	0.0	$	&	$	2.8	^{	+1.0	}_{	-0.6	}	$	&	$	9.2	^{	+1.3	}_{	-0.8	}	\pm	0.5	$	&	$	10	$	\\
	900		&	$	150	\pm	0.9	$	&	$	4.0	\pm	0.3	$	&	$	0.0	^{	+1.2	}_{	-0.0	}	$	&	$	0.0	\pm	0.0	$	&	$	2.6	^{	+0.8	}_{	-0.5	}	$	&	$	6.6	^{	+1.4	}_{	-0.6	}	\pm	0.4	$	&	$	8	$	\\
	950		&	$	110	\pm	0.6	$	&	$	3.6	\pm	0.5	$	&	$	0.0	^{	+0.9	}_{	-0.0	}	$	&	$	0.0	\pm	0.0	$	&	$	2.1	^{	+0.7	}_{	-0.5	}	$	&	$	5.7	^{	+1.3	}_{	-0.7	}	\pm	0.3	$	&	$	5	$	\\
	1000		&	$	77	\pm	0.4	$	&	$	2.2	\pm	0.1	$	&	$	0.0	^{	+0.7	}_{	-0.0	}	$	&	$	0.0	\pm	0.0	$	&	$	1.9	^{	+0.7	}_{	-0.4	}	$	&	$	4.1	^{	+1.0	}_{	-0.5	}	\pm	0.2	$	&	$	5	$	\\
	1050		&	$	55	\pm	0.3	$	&	$	1.8	\pm	0.1	$	&	$	0.0	^{	+0.3	}_{	-0.0	}	$	&	$	0.0	\pm	0.0	$	&	$	1.4	^{	+0.6	}_{	-0.4	}	$	&	$	3.2	^{	+0.7	}_{	-0.4	}	\pm	0.2	$	&	$	4	$	\\
	1100		&	$	41	\pm	0.2	$	&	$	1.8	\pm	0.1	$	&	$	0.0	^{	+0.3	}_{	-0.0	}	$	&	$	0.0	\pm	0.0	$	&	$	1.4	^{	+0.6	}_{	-0.4	}	$	&	$	3.2	^{	+0.7	}_{	-0.4	}	\pm	0.2	$	&	$	4	$	\\
	1150		&	$	31	\pm	0.2	$	&	$	1.8	\pm	0.1	$	&	$	0.0	^{	+0.3	}_{	-0.0	}	$	&	$	0.0	\pm	0.0	$	&	$	1.4	^{	+0.6	}_{	-0.4	}	$	&	$	3.2	^{	+0.7	}_{	-0.4	}	\pm	0.2	$	&	$	4	$	\\
	1200		&	$	23	\pm	0.1	$	&	$	1.8	\pm	0.1	$	&	$	0.0	^{	+0.3	}_{	-0.0	}	$	&	$	0.0	\pm	0.0	$	&	$	1.4	^{	+0.6	}_{	-0.4	}	$	&	$	3.2	^{	+0.7	}_{	-0.4	}	\pm	0.2	$	&	$	4	$	\\
	1250		&	$	17	\pm	0.1	$	&	$	1.8	\pm	0.1	$	&	$	0.0	^{	+0.3	}_{	-0.0	}	$	&	$	0.0	\pm	0.0	$	&	$	1.4	^{	+0.6	}_{	-0.4	}	$	&	$	3.2	^{	+0.7	}_{	-0.4	}	\pm	0.2	$	&	$	4	$	\\
	1300		&	$	13	\pm	0.1	$	&	$	1.8	\pm	0.1	$	&	$	0.0	^{	+0.3	}_{	-0.0	}	$	&	$	0.0	\pm	0.0	$	&	$	1.4	^{	+0.6	}_{	-0.4	}	$	&	$	3.2	^{	+0.7	}_{	-0.4	}	\pm	0.2	$	&	$	4	$	\\
	1350		&	$	9.8	\pm	0.0	$	&	$	1.8	\pm	0.1	$	&	$	0.0	^{	+0.3	}_{	-0.0	}	$	&	$	0.0	\pm	0.0	$	&	$	1.4	^{	+0.6	}_{	-0.4	}	$	&	$	3.2	^{	+0.7	}_{	-0.4	}	\pm	0.2	$	&	$	4	$	\\
	1400		&	$	7.4	\pm	0.0	$	&	$	1.8	\pm	0.1	$	&	$	0.0	^{	+0.3	}_{	-0.0	}	$	&	$	0.0	\pm	0.0	$	&	$	1.4	^{	+0.6	}_{	-0.4	}	$	&	$	3.2	^{	+0.7	}_{	-0.4	}	\pm	0.2	$	&	$	4	$	\\
	1450		&	$	5.6	\pm	0.0	$	&	$	1.8	\pm	0.1	$	&	$	0.0	^{	+0.3	}_{	-0.0	}	$	&	$	0.0	\pm	0.0	$	&	$	1.4	^{	+0.6	}_{	-0.4	}	$	&	$	3.2	^{	+0.7	}_{	-0.4	}	\pm	0.2	$	&	$	4	$	\\
	1500		&	$	4.2	\pm	0.0	$	&	$	1.8	\pm	0.1	$	&	$	0.0	^{	+0.3	}_{	-0.0	}	$	&	$	0.0	\pm	0.0	$	&	$	1.4	^{	+0.6	}_{	-0.4	}	$	&	$	3.2	^{	+0.7	}_{	-0.4	}	\pm	0.2	$	&	$	4	$	\\
	1550		&	$	3.2	\pm	0.0	$	&	$	1.8	\pm	0.1	$	&	$	0.0	^{	+0.3	}_{	-0.0	}	$	&	$	0.0	\pm	0.0	$	&	$	1.4	^{	+0.6	}_{	-0.4	}	$	&	$	3.2	^{	+0.7	}_{	-0.4	}	\pm	0.2	$	&	$	4	$	\\
	1600		&	$	2.4	\pm	0.0	$	&	$	1.8	\pm	0.1	$	&	$	0.0	^{	+0.3	}_{	-0.0	}	$	&	$	0.0	\pm	0.0	$	&	$	1.4	^{	+0.6	}_{	-0.4	}	$	&	$	3.2	^{	+0.7	}_{	-0.4	}	\pm	0.2	$	&	$	4	$	\\
	1650		&	$	1.8	\pm	0.0	$	&	$	1.8	\pm	0.1	$	&	$	0.0	^{	+0.3	}_{	-0.0	}	$	&	$	0.0	\pm	0.0	$	&	$	1.4	^{	+0.6	}_{	-0.4	}	$	&	$	3.2	^{	+0.7	}_{	-0.4	}	\pm	0.2	$	&	$	4	$	\\
	1700		&	$	1.4	\pm	0.0	$	&	$	1.8	\pm	0.1	$	&	$	0.0	^{	+0.3	}_{	-0.0	}	$	&	$	0.0	\pm	0.0	$	&	$	1.4	^{	+0.6	}_{	-0.4	}	$	&	$	3.2	^{	+0.7	}_{	-0.4	}	\pm	0.2	$	&	$	4	$	\\
	1750		&	$	1.1	\pm	0.0	$	&	$	1.8	\pm	0.1	$	&	$	0.0	^{	+0.3	}_{	-0.0	}	$	&	$	0.0	\pm	0.0	$	&	$	1.4	^{	+0.6	}_{	-0.4	}	$	&	$	3.2	^{	+0.7	}_{	-0.4	}	\pm	0.2	$	&	$	4	$	\\
	1800		&	$	0.8	\pm	0.0	$	&	$	1.8	\pm	0.1	$	&	$	0.0	^{	+0.3	}_{	-0.0	}	$	&	$	0.0	\pm	0.0	$	&	$	1.4	^{	+0.6	}_{	-0.4	}	$	&	$	3.2	^{	+0.7	}_{	-0.4	}	\pm	0.2	$	&	$	4	$	\\
	1850		&	$	0.6	\pm	0.0	$	&	$	1.8	\pm	0.1	$	&	$	0.0	^{	+0.3	}_{	-0.0	}	$	&	$	0.0	\pm	0.0	$	&	$	1.4	^{	+0.6	}_{	-0.4	}	$	&	$	3.2	^{	+0.7	}_{	-0.4	}	\pm	0.2	$	&	$	4	$	\\
	1900		&	$	0.5	\pm	0.0	$	&	$	1.8	\pm	0.1	$	&	$	0.0	^{	+0.3	}_{	-0.0	}	$	&	$	0.0	\pm	0.0	$	&	$	1.4	^{	+0.6	}_{	-0.4	}	$	&	$	3.2	^{	+0.7	}_{	-0.4	}	\pm	0.2	$	&	$	4	$	\\
	1950		&	$	0.4	\pm	0.0	$	&	$	1.8	\pm	0.1	$	&	$	0.0	^{	+0.3	}_{	-0.0	}	$	&	$	0.0	\pm	0.0	$	&	$	1.4	^{	+0.6	}_{	-0.4	}	$	&	$	3.2	^{	+0.7	}_{	-0.4	}	\pm	0.2	$	&	$	4	$	\\
	2000		&	$	0.3	\pm	0.0	$	&	$	1.8	\pm	0.1	$	&	$	0.0	^{	+0.3	}_{	-0.0	}	$	&	$	0.0	\pm	0.0	$	&	$	1.4	^{	+0.6	}_{	-0.4	}	$	&	$	3.2	^{	+0.7	}_{	-0.4	}	\pm	0.2	$	&	$	4	$	\\
    \end{scotch}
}
\end{table*}

\begin{table*}[htpb]
  \centering
    \topcaption{Event yields after the optimized \enujj\ selections.
    Uncertainties are statistical except for the total background, where both
    statistical and systematic uncertainties are shown. An entry of $0.0$ quoted
    for the uncertainty indicates that its value is negligibly small. LQ masses
    are given in units of \GeV and init. sel. refers to initial selection.}
    \label{tab:enujjFinalSels}
    \def\arraystretch{1.5}
    \cmsTable{
    \begin{scotch}{cccccccc}

	LQ mass		&		Signal				&		\wjets				&		\ttbar							&		Multijet				&		\vv, \PZ, single \cPqt, \gammajets							&		Total background									&		Data		\\ \hline
init. sel.		&	$	\NA			$	&	$	47900	\pm	160	$	&	$	66900	\pm	110				$	&	$	2800	\pm	15	$	&	$	11300	\pm	72				$	&	$	128900	\pm	210				\pm	8800	$	&	$	125076	$	\\
\noalign{\vskip 2mm}
	200		&	$	130800	\pm	1600	$	&	$	40100	\pm	150	$	&	$	52800	\pm	94				$	&	$	2100	\pm	11	$	&	$	9600	\pm	57				$	&	$	104500	\pm	190				\pm	7300	$	&	$	101618	$	\\
	250		&	$	44200	\pm	520	$	&	$	1800	\pm	25	$	&	$	3800	\pm	25				$	&	$	300	\pm	2.3	$	&	$	1300	\pm	38				$	&	$	7100	\pm	52				\pm	430	$	&	$	7151	$	\\
	300		&	$	19800	\pm	220	$	&	$	800	\pm	15	$	&	$	1400	\pm	16				$	&	$	120	\pm	1.4	$	&	$	660	\pm	37				$	&	$	3000	\pm	43				\pm	170	$	&	$	3164	$	\\
	350		&	$	9800	\pm	100	$	&	$	410	\pm	13	$	&	$	610	\pm	10				$	&	$	62	\pm	1.0	$	&	$	330	\pm	11				$	&	$	1400	\pm	20				\pm	88	$	&	$	1539	$	\\
	400		&	$	5100	\pm	51	$	&	$	230	\pm	8.9	$	&	$	300	\pm	7.2				$	&	$	37	\pm	0.8	$	&	$	200	\pm	10				$	&	$	760	\pm	15				\pm	74	$	&	$	847	$	\\
	450		&	$	2900	\pm	27	$	&	$	150	\pm	6.0	$	&	$	160	\pm	5.2				$	&	$	28	\pm	0.8	$	&	$	120	\pm	9.6				$	&	$	460	\pm	12				\pm	31	$	&	$	496	$	\\
	500		&	$	1700	\pm	15	$	&	$	90	\pm	4.1	$	&	$	88	\pm	3.9				$	&	$	21	\pm	0.8	$	&	$	75	^{	+3.9	}_{	-3.3	}	$	&	$	270	^{	+6.9	}_{	-6.6	}	\pm	21	$	&	$	298	$	\\
	550		&	$	990	\pm	8.8	$	&	$	59	\pm	5.2	$	&	$	49	\pm	2.9				$	&	$	9.1	\pm	0.4	$	&	$	53	^{	+3.5	}_{	-2.9	}	$	&	$	170	^{	+6.9	}_{	-6.6	}	\pm	13	$	&	$	195	$	\\
	600		&	$	620	\pm	5.3	$	&	$	45	\pm	5.1	$	&	$	32	\pm	2.3				$	&	$	6.1	\pm	0.4	$	&	$	36	^{	+2.8	}_{	-2.2	}	$	&	$	120	^{	+6.3	}_{	-6.0	}	\pm	12	$	&	$	132	$	\\
	650		&	$	400	\pm	3.3	$	&	$	34	\pm	5.0	$	&	$	20	\pm	1.8				$	&	$	5.0	\pm	0.4	$	&	$	26	^{	+2.5	}_{	-1.9	}	$	&	$	84	^{	+5.9	}_{	-5.7	}	\pm	8.1	$	&	$	94	$	\\
	700		&	$	270	\pm	2.1	$	&	$	22	\pm	1.2	$	&	$	12	\pm	1.5				$	&	$	4.2	\pm	0.5	$	&	$	18	^{	+2.1	}_{	-1.5	}	$	&	$	56	^{	+2.9	}_{	-2.5	}	\pm	6.1	$	&	$	71	$	\\
	750		&	$	180	\pm	1.4	$	&	$	15	\pm	0.9	$	&	$	10	\pm	1.3				$	&	$	3.7	\pm	0.5	$	&	$	13	^{	+2.1	}_{	-1.3	}	$	&	$	42	^{	+2.7	}_{	-2.1	}	\pm	4.9	$	&	$	49	$	\\
	800		&	$	130	\pm	0.9	$	&	$	13	\pm	1.0	$	&	$	6.3	\pm	1.0				$	&	$	3.4	\pm	0.6	$	&	$	9.8	^{	+2.0	}_{	-1.1	}	$	&	$	32	^{	+2.5	}_{	-1.9	}	\pm	4.6	$	&	$	38	$	\\
	850		&	$	86	\pm	0.6	$	&	$	13	\pm	1.1	$	&	$	5.2	\pm	0.9				$	&	$	3.2	\pm	0.7	$	&	$	7.0	^{	+2.0	}_{	-1.2	}	$	&	$	28	^{	+2.6	}_{	-2.0	}	\pm	4.8	$	&	$	28	$	\\
	900		&	$	61	\pm	0.4	$	&	$	11	\pm	1.2	$	&	$	3.8	\pm	0.8				$	&	$	3.0	\pm	0.7	$	&	$	6.3	^{	+2.0	}_{	-1.1	}	$	&	$	24	^{	+2.6	}_{	-2.0	}	\pm	4.1	$	&	$	21	$	\\
	950		&	$	44	\pm	0.3	$	&	$	8.4	\pm	1.0	$	&	$	3.0	\pm	0.7				$	&	$	0.7	\pm	0.1	$	&	$	5.7	^{	+2.0	}_{	-1.1	}	$	&	$	18	^{	+2.3	}_{	-1.6	}	\pm	3.3	$	&	$	20	$	\\
	1000		&	$	31	\pm	0.2	$	&	$	7.9	\pm	0.9	$	&	$	2.2	\pm	0.6				$	&	$	0.6	\pm	0.1	$	&	$	4.8	^{	+2.0	}_{	-1.1	}	$	&	$	16	^{	+2.3	}_{	-1.5	}	\pm	2.8	$	&	$	15	$	\\
	1050		&	$	23	\pm	0.2	$	&	$	7.1	\pm	0.9	$	&	$	1.4	^{	+0.7	}_{	-0.5	}	$	&	$	0.5	\pm	0.1	$	&	$	4.4	^{	+2.0	}_{	-1.1	}	$	&	$	13	^{	+2.3	}_{	-1.4	}	\pm	2.5	$	&	$	14	$	\\
	1100		&	$	17	\pm	0.1	$	&	$	5.9	\pm	0.8	$	&	$	1.2	^{	+0.6	}_{	-0.4	}	$	&	$	0.5	\pm	0.1	$	&	$	4.0	^{	+2.0	}_{	-1.0	}	$	&	$	12	^{	+2.3	}_{	-1.4	}	\pm	2.1	$	&	$	12	$	\\
	1150		&	$	12	\pm	0.1	$	&	$	5.4	\pm	0.9	$	&	$	0.9	^{	+0.6	}_{	-0.4	}	$	&	$	0.4	\pm	0.1	$	&	$	3.3	^{	+2.0	}_{	-1.0	}	$	&	$	10	^{	+2.3	}_{	-1.4	}	\pm	1.7	$	&	$	12	$	\\
	1200		&	$	9.1	\pm	0.1	$	&	$	5.2	\pm	1.1	$	&	$	0.7	^{	+0.6	}_{	-0.4	}	$	&	$	0.4	\pm	0.1	$	&	$	3.2	^{	+2.0	}_{	-1.0	}	$	&	$	9.5	^{	+2.3	}_{	-1.5	}	\pm	1.6	$	&	$	10	$	\\
	1250		&	$	7.1	\pm	0.0	$	&	$	5.0	\pm	1.1	$	&	$	0.7	^{	+0.6	}_{	-0.4	}	$	&	$	0.4	\pm	0.1	$	&	$	3.0	^{	+2.0	}_{	-1.0	}	$	&	$	9.1	^{	+2.3	}_{	-1.5	}	\pm	1.5	$	&	$	9	$	\\
	1300		&	$	5.4	\pm	0.0	$	&	$	5.0	\pm	1.1	$	&	$	0.7	^{	+0.6	}_{	-0.4	}	$	&	$	0.4	\pm	0.1	$	&	$	3.0	^{	+2.0	}_{	-1.0	}	$	&	$	9.1	^{	+2.3	}_{	-1.5	}	\pm	1.5	$	&	$	9	$	\\
	1350		&	$	4.1	\pm	0.0	$	&	$	5.0	\pm	1.1	$	&	$	0.7	^{	+0.6	}_{	-0.4	}	$	&	$	0.4	\pm	0.1	$	&	$	3.0	^{	+2.0	}_{	-1.0	}	$	&	$	9.1	^{	+2.3	}_{	-1.5	}	\pm	1.5	$	&	$	9	$	\\
	1400		&	$	3.1	\pm	0.0	$	&	$	5.0	\pm	1.1	$	&	$	0.7	^{	+0.6	}_{	-0.4	}	$	&	$	0.4	\pm	0.1	$	&	$	3.0	^{	+2.0	}_{	-1.0	}	$	&	$	9.1	^{	+2.3	}_{	-1.5	}	\pm	1.5	$	&	$	9	$	\\
	1450		&	$	2.4	\pm	0.0	$	&	$	5.0	\pm	1.1	$	&	$	0.7	^{	+0.6	}_{	-0.4	}	$	&	$	0.4	\pm	0.1	$	&	$	3.0	^{	+2.0	}_{	-1.0	}	$	&	$	9.1	^{	+2.3	}_{	-1.5	}	\pm	1.5	$	&	$	9	$	\\
	1500		&	$	1.9	\pm	0.0	$	&	$	5.0	\pm	1.1	$	&	$	0.7	^{	+0.6	}_{	-0.4	}	$	&	$	0.4	\pm	0.1	$	&	$	3.0	^{	+2.0	}_{	-1.0	}	$	&	$	9.1	^{	+2.3	}_{	-1.5	}	\pm	1.5	$	&	$	9	$	\\
	1550		&	$	1.4	\pm	0.0	$	&	$	5.0	\pm	1.1	$	&	$	0.7	^{	+0.6	}_{	-0.4	}	$	&	$	0.4	\pm	0.1	$	&	$	3.0	^{	+2.0	}_{	-1.0	}	$	&	$	9.1	^{	+2.3	}_{	-1.5	}	\pm	1.5	$	&	$	9	$	\\
	1600		&	$	1.1	\pm	0.0	$	&	$	5.0	\pm	1.1	$	&	$	0.7	^{	+0.6	}_{	-0.4	}	$	&	$	0.4	\pm	0.1	$	&	$	3.0	^{	+2.0	}_{	-1.0	}	$	&	$	9.1	^{	+2.3	}_{	-1.5	}	\pm	1.5	$	&	$	9	$	\\
	1650		&	$	0.8	\pm	0.0	$	&	$	5.0	\pm	1.1	$	&	$	0.7	^{	+0.6	}_{	-0.4	}	$	&	$	0.4	\pm	0.1	$	&	$	3.0	^{	+2.0	}_{	-1.0	}	$	&	$	9.1	^{	+2.3	}_{	-1.5	}	\pm	1.5	$	&	$	9	$	\\
	1700		&	$	0.6	\pm	0.0	$	&	$	5.0	\pm	1.1	$	&	$	0.7	^{	+0.6	}_{	-0.4	}	$	&	$	0.4	\pm	0.1	$	&	$	3.0	^{	+2.0	}_{	-1.0	}	$	&	$	9.1	^{	+2.3	}_{	-1.5	}	\pm	1.5	$	&	$	9	$	\\
	1750		&	$	0.5	\pm	0.0	$	&	$	5.0	\pm	1.1	$	&	$	0.7	^{	+0.6	}_{	-0.4	}	$	&	$	0.4	\pm	0.1	$	&	$	3.0	^{	+2.0	}_{	-1.0	}	$	&	$	9.1	^{	+2.3	}_{	-1.5	}	\pm	1.5	$	&	$	9	$	\\
	1800		&	$	0.4	\pm	0.0	$	&	$	5.0	\pm	1.1	$	&	$	0.7	^{	+0.6	}_{	-0.4	}	$	&	$	0.4	\pm	0.1	$	&	$	3.0	^{	+2.0	}_{	-1.0	}	$	&	$	9.1	^{	+2.3	}_{	-1.5	}	\pm	1.5	$	&	$	9	$	\\
	1850		&	$	0.3	\pm	0.0	$	&	$	5.0	\pm	1.1	$	&	$	0.7	^{	+0.6	}_{	-0.4	}	$	&	$	0.4	\pm	0.1	$	&	$	3.0	^{	+2.0	}_{	-1.0	}	$	&	$	9.1	^{	+2.3	}_{	-1.5	}	\pm	1.5	$	&	$	9	$	\\
	1900		&	$	0.2	\pm	0.0	$	&	$	5.0	\pm	1.1	$	&	$	0.7	^{	+0.6	}_{	-0.4	}	$	&	$	0.4	\pm	0.1	$	&	$	3.0	^{	+2.0	}_{	-1.0	}	$	&	$	9.1	^{	+2.3	}_{	-1.5	}	\pm	1.5	$	&	$	9	$	\\
	1950		&	$	0.2	\pm	0.0	$	&	$	5.0	\pm	1.1	$	&	$	0.7	^{	+0.6	}_{	-0.4	}	$	&	$	0.4	\pm	0.1	$	&	$	3.0	^{	+2.0	}_{	-1.0	}	$	&	$	9.1	^{	+2.3	}_{	-1.5	}	\pm	1.5	$	&	$	9	$	\\
	2000		&	$	0.1	\pm	0.0	$	&	$	5.0	\pm	1.1	$	&	$	0.7	^{	+0.6	}_{	-0.4	}	$	&	$	0.4	\pm	0.1	$	&	$	3.0	^{	+2.0	}_{	-1.0	}	$	&	$	9.1	^{	+2.3	}_{	-1.5	}	\pm	1.5	$	&	$	9	$	\\
    \end{scotch}
    }
\end{table*}
\cleardoublepage \section{The CMS Collaboration \label{app:collab}}\begin{sloppypar}\hyphenpenalty=5000\widowpenalty=500\clubpenalty=5000\input{EXO-17-009-authorlist.tex}\end{sloppypar}
\end{document}

%% file: EXO-17-009-authorlist.tex
\vskip\cmsinstskip
\textbf{Yerevan Physics Institute, Yerevan, Armenia}\\*[0pt]
A.M.~Sirunyan, A.~Tumasyan
\vskip\cmsinstskip
\textbf{Institut f\"{u}r Hochenergiephysik, Wien, Austria}\\*[0pt]
W.~Adam, F.~Ambrogi, E.~Asilar, T.~Bergauer, J.~Brandstetter, M.~Dragicevic, J.~Er\"{o}, A.~Escalante~Del~Valle, M.~Flechl, R.~Fr\"{u}hwirth\cmsAuthorMark{1}, V.M.~Ghete, J.~Hrubec, M.~Jeitler\cmsAuthorMark{1}, N.~Krammer, I.~Kr\"{a}tschmer, D.~Liko, T.~Madlener, I.~Mikulec, N.~Rad, H.~Rohringer, J.~Schieck\cmsAuthorMark{1}, R.~Sch\"{o}fbeck, M.~Spanring, D.~Spitzbart, A.~Taurok, W.~Waltenberger, J.~Wittmann, C.-E.~Wulz\cmsAuthorMark{1}, M.~Zarucki
% ...
\vskip\cmsinstskip
% ...
1:  Also at Vienna University of Technology, Vienna, Austria\\
% ...
